\pdfoutput=1
\documentclass[aps,prd,amsmath,floats,floatfix, twocolumn,
superscriptaddress,nofootinbib,showpacs]{revtex4-2}

\usepackage[T1]{fontenc}
\usepackage[utf8]{inputenc}
\usepackage{lmodern}
\usepackage{verbatim}
\usepackage{widetable}

\usepackage[dvipsnames, usenames]{xcolor}
\definecolor{linkcolor}{rgb}{0.0,0.3,0.5}
\usepackage[hypertexnames=false, unicode, colorlinks=true, linkcolor=linkcolor,
citecolor=linkcolor, filecolor=linkcolor,urlcolor=linkcolor,
pdfusetitle]{hyperref}
\usepackage{orcidlink}

\usepackage[all]{hypcap}
\usepackage{graphicx}
\usepackage{xspace}

\usepackage{amssymb}
\usepackage[normalem]{ulem} 
\usepackage{bm} 
\usepackage{enumitem,amssymb}

\usepackage[caption=false]{subfig}

\usepackage[justification=raggedright,singlelinecheck=false,font=small]{caption}
\usepackage{orcidlink}

\begin{document}

\title{Complex frequency evolution of direct waves from binary black hole mergers}
\author{Ling Sun\orcidlink{0000-0001-7959-892X}}
\affiliation{OzGrav-ANU, Centre for Gravitational Astrophysics, Research School of Physics and Research School of Astronomy \& Astrophysics, The Australian National University, ACT 2601, Australia}

\author{Sizheng Ma\orcidlink{0000-0002-4645-453X}}
\affiliation{Perimeter Institute for Theoretical Physics, Waterloo, ON N2L2Y5, Canada}

\author{Patrick P. Chen\orcidlink{0009-0004-7467-8786}}
\affiliation{Burke Institute for Theoretical Physics and Theoretical Astrophysics 350-17, California Institute of Technology, Pasadena, CA 91125, USA}
\affiliation{University of Cambridge, Cambridge CB2 1TN, United Kingdom}

\author{Andrew Laeuger\orcidlink{0000-0002-8212-6496}}
\affiliation{Burke Institute for Theoretical Physics and Theoretical Astrophysics 350-17, California Institute of Technology, Pasadena, CA 91125, USA}

\author{Neil Lu\orcidlink{0000-0002-8861-9902}}
\affiliation{OzGrav-ANU, Centre for Gravitational Astrophysics, Research School of Physics and Research School of Astronomy \& Astrophysics, The Australian National University, ACT 2601, Australia}

\author{Yanbei Chen\orcidlink{0000-0002-9730-9463}}
\affiliation{Burke Institute for Theoretical Physics and Theoretical Astrophysics 350-17, California Institute of Technology, Pasadena, CA 91125, USA}


\begin{abstract}
While no signal originating at a black hole's event horizon can reach future null infinity, information about the horizon and its immediate vicinity can be encoded in asymptotic properties of waves emitted by matter or field perturbations falling toward a growing/forming horizon. 
Within the response-filtered framework, we use the term ``direct wave'' to denote source-sensitive information associated with the plunge and remnant formation that is revealed by filtering the black-hole response.
Unlike a stationary damped sinusoid with a fixed complex frequency, the direct wave is determined by the evolving source dynamics and is characterized by an evolving instantaneous complex frequency whose late-time behavior is tied to the angular velocity $\Omega_H$ and surface gravity $\kappa_H$ of the remnant horizon. 
We study the complex-frequency evolution of the direct wave by removing quasinormal modes from numerical-relativity waveforms using rational filters. 
Across the systems considered, the complex-frequency trajectories of direct waves---before numerical contamination sets in---exhibit a systematic dependence on the remnant spin: the real frequency evolves toward $2\Omega_H$ from above for lower spins ($\chi_f \lesssim 0.7$) and from below for higher spins ($\chi_f \gtrsim 0.7$), while the instantaneous decay rate generally increases, with best-resolved cases approaching the expected source decay rate of $3\kappa_H$. 
These results resemble those obtained previously for particle plunges and are consistent with a late-time approach to horizon-controlled source asymptotics, as frame dragging and gravitational redshift increasingly affect the near-horizon dynamics. 
A special case occurs near $\chi_f\sim0.7$, characteristic of the remnant of an equal-mass, non-spinning binary, for which the early-time real frequency is already close to $2\Omega_H$ because the orbital frequency associated with the binary motion transitions smoothly to the angular velocity of the remnant horizon.
Finite-time deviations from the horizon-controlled limit therefore do not undermine the connection between the direct wave and the remnant horizon, but instead carry information about merger/collapse dynamics. Fully characterizing the direct-wave component requires benchmarking its complete evolution against numerical-relativity predictions. 
We further show that approximate pole--zero pairing in the Kerr response provides a physical basis for a minimal filter set that sufficiently suppresses the quasinormal-mode features of the black-hole response while revealing the source trajectory information encoded in the direct wave. 
\end{abstract}

\maketitle

\section{Introduction}

Recent theoretical and observational studies of the \textit{direct wave}~\cite{Oshita:2025qmn,Lu2026} have generated substantial interest. These studies suggest that a portion of the merger--ringdown gravitational-wave signal from binary black hole mergers may encode information associated with the plunge, remnant formation, and the horizon of the remnant black hole. The presence of this component has recently been independently supported using a different methodology~\cite{dyer2026}, and the direct wave has also been proposed as a test of Hawking's area law~\cite{chung2026}. Subsequent work has sharpened several questions concerning its definition, frequency evolution, and connection to the remnant horizon. Ref.~\cite{kankani2026} questioned the interpretation advanced in Refs.~\cite{Oshita:2025qmn,Lu2026}, as well as its application in Ref.~\cite{chung2026}, on the grounds that the direct wave had effectively been treated as having a fixed frequency. At the same time, Ref.~\cite{kankani2026} emphasized that the direct-wave frequency evolves with time and interpreted the discrepancy between its early-time value and the horizon rotation frequency as evidence against a connection to the horizon. On the theoretical side, Ref.~\cite{kuntz2026} argued that the direct wave should vanish identically, at least for a Schwarzschild black hole. Together, these developments motivate a precise definition of the direct wave and a clearer distinction between its evolving behavior at finite times and its horizon-controlled late-time asymptote.

It is therefore useful to first clarify what the direct wave means within the filtered-waveform framework. In Refs.~\cite{Oshita:2025qmn,Lu2026}, the term refers operationally to the decaying cycles that remain at the end of the waveform after applying rational filters~\cite{Ma:2022wpv,Ma:2023cwe,Ma:2023vvr}. More generally, we identify its imprint as the structured near-merger signal revealed by a specified response-informed filter. Physically, we use the term ``direct wave'' to refer to source-driven information associated with the plunge and remnant formation, as transmitted and screened by the black-hole response. From the standard perspective of black-hole perturbation theory, the waveform is conventionally decomposed into the prompt response, quasinormal-mode (QNM) contribution, and late-time tail~\cite{Leaver:1986gd,Andersson:1996cm,Price:1971fb,BertiReview:2009}. Within the present framework, we do not regard the direct wave as an additional component of this decomposition. Rather, response-informed filtering reorganizes the same waveform in the time domain so as to expose information about the source trajectory that is otherwise mixed with the resonant black-hole response~\cite{ma2026}. 

Direct waves are {\it continuously driven} by matter or spacetime perturbations that fall into a black hole, whose asymptotic behavior is intimately tied to properties of the horizon. The extraction of such source-driven information can be traced back to the earlier proposal of a ``horizon mode.'' Mino and Brink~\cite{Mino:2008at} argued that radiation from a particle plunging toward a Kerr horizon should approach, at late times, a complex frequency $\omega_H=m\Omega_H-i\kappa_H$, where $\Omega_H$ is the horizon angular velocity, $\kappa_H$ is the surface gravity, and $m$ is the azimuthal quantum number of the radiation. Zimmerman and Chen~\cite{Zimmerman:2011dx} subsequently corrected the predicted late-time decay rate and argued that the relevant complex frequency should instead approach $\omega_H=m\Omega_H-2i\kappa_H$. They also emphasized, however, that {\it stationary} components at the characteristic horizon frequencies $m\Omega_H-in\kappa_H$ are screened by the black-hole potential barrier and therefore cannot be observed at infinity. Much later, Oshita {\it et al.}~\cite{Oshita:2025qmn} recognized that radiation emitted by a plunging source with a {\it dynamically evolving} frequency need not be completely screened: the evolving signal can partially transmit through the potential barrier and produce an observable direct wave. Lu {\it et al.}~\cite{Lu2026} subsequently applied this framework to the binary black hole merger GW250114~\cite{GW250114,GW250114com}.

The dynamical evolution of the direct-wave complex frequency is an essential part of the interpretation developed in Refs.~\cite{Oshita:2025qmn,Lu2026}. Oshita {\it et al.}~\cite{Oshita:2025qmn} explicitly modeled the evolution of the direct-wave complex frequency under an analytical approximation. Lu {\it et al.}~\cite{Lu2026} performed a matched-filter analysis based on this analytical model 
with an evolving complex frequency, and also measured the effective direct-wave frequency and decay rate over several consecutive time windows using a model-agnostic approach. A fixed-frequency model of the direct wave, if carefully calibrated against numerical simulations as motivated by Ref.~\cite{chung2026}, can provide meaningful measurements of remnant parameters. Such a model, however, differs from the {\it dynamical} framework developed by Oshita {\it et al.} and applied by Lu {\it et al.}, in which the frequency evolution is an essential part of the signal and reflects the dynamical formation of the remnant black hole. A discrepancy between the instantaneous direct-wave frequency and the asymptotic horizon frequency at any particular time therefore does not, by itself, provide strong evidence against a connection to the horizon. Rather, the evolution of the complex frequency is itself part of the physical signal: it encodes how the observed radiation progressively acquires the characteristic imprint of the near-horizon dynamics as the source approaches the horizon. Thus, deviations from the horizon frequency and the dependence of an effective damping rate on a finite-time fitting interval, as discussed by Ref.~\cite{Oshita:2025qmn} and observed by Ref.~\cite{Lu2026}, do not contradict the connection between direct waves and the black hole horizon.

In this work, rather than attempting to provide a comprehensive theory of the direct wave, its relation to merger dynamics, and its connection to the remnant black hole properties, we aim to clarify this main physical picture. To illustrate the distinction between the stationary and dynamical pictures, we track the instantaneous complex-frequency evolution of the direct wave in a series of numerical-relativity (NR) waveforms, including those obtained from conventional extrapolation and from the newer Cauchy--Characteristic-Extraction (CCE) method~\cite{Moxon:2020gha,Moxon:2021gbv}. To extract the direct waves, we remove the QNMs from these waveforms using rational filters~\cite{Ma:2022wpv,Ma:2023cwe,Ma:2023vvr}. Over the interval in which the direct-wave component can be reliably extracted numerically, we find that its complex frequency follows remnant-spin-dependent trajectories with horizon-controlled asymptotics: the real part approaches $m\Omega_H$ from above or below, depending on whether the remnant spin $\chi_f$ is below or above $\sim0.7$, while the instantaneous decay rate increases, with best-resolved cases reaching values near the expected source decay rate of $3\kappa_H$. These results for comparable-mass binaries closely resemble those obtained by Ref.~\cite{Oshita:2025qmn} for a particle plunging into Kerr.

We further show that the pole--zero structure of the black-hole response provides a physical basis for a minimal filter set that sufficiently suppresses the QNM features of the response while revealing the source-trajectory information encoded in the direct wave. The approximate pole--zero pairing present for moderate and high Kerr spins breaks down in the Schwarzschild limit, helping to explain the qualitatively different behavior found in Refs.~\cite{ma2026,kuntz2026}. Our results therefore further support the physical interpretation advanced by Oshita {\it et al.}~\cite{Oshita:2025qmn} and Lu {\it et al.}~\cite{Lu2026}, while clarifying both the distinction between an evolving direct wave at finite times and a stationary mode with a fixed complex frequency and the role of the black-hole response in revealing the source-driven signal.

\section{Physical motivation for the direct wave}

In this section, we review the physical picture underlying the direct-wave framework, from the screening of stationary horizon frequencies to the observable response driven by a dynamically plunging source, and extend this picture from particle plunges to comparable-mass binary mergers.

\subsection{Horizon mode and Matsubara frequencies}

Mino and Brink~\cite{Mino:2008at} argued that a particle plunging toward the horizon of a Kerr black hole must, as seen by a distant observer, rotate with an angular frequency approaching $\Omega_H$, and therefore emit gravitational waves at frequencies approaching $m\Omega_H$. They also obtained an imaginary part $-i\kappa_H$, indicating the damping rate.
Zimmerman and Chen~\cite{Zimmerman:2011dx} subsequently identified a typographical error in the analysis of Mino and Brink~\cite{Mino:2008at} and argued that the relevant complex frequency should instead approach
\begin{equation}
    \omega \rightarrow m\Omega_H-2i\kappa_H ,
\end{equation}
or decay more rapidly. The decay rate $2\kappa_H$ was inferred by imposing regularity and smoothness conditions on curvature perturbations at the surface of the collapsing object, evaluated in the local inertial frames at the horizon.
More recent studies find that the leading late-time effective-source frequency approaches $m\Omega_H-3i\kappa_H$~\cite{weller2026,kubota2026}.

Note that both Refs.~\cite{Mino:2008at,Zimmerman:2011dx} proposed modes with stationary complex frequencies, which may have contributed to some of the subsequent confusion. 
However, stationary components at Matsubara frequencies~\cite{Matsubara1955} associated with the horizon,
\begin{equation}
    \omega_{\rm MB}^{(j)}=m\Omega_H-i j \kappa_H,
    \qquad j=1,2,3,\ldots,
\end{equation}
are screened by the black-hole potential barrier and cannot escape to infinity. 
This can be understood both from the thermodynamic interpretation~\cite{Hawking:1975vcx,Birmingham:2001pj,Son:2002sd} and from the vanishing transmission coefficients at these frequencies derived by Mano, Suzuki, and Takasugi~\cite{Mano:1996vt,SasakiReview}.
The existence of such characteristic frequencies near the horizon therefore does not, by itself, imply that radiation at these frequencies is observable at infinity. The original horizon-mode argument was thus incomplete as a motivation for an observable signature. This issue was recognized by Zimmerman, Mark, and Chen~\cite{Zimmerman:2018APS} and has recently been emphasized by Kuntz and Della Rocca~\cite{kuntz2026}.

More specifically, when computing the Teukolsky function $\psi_4$ emitted by a particle plunging into a Kerr black hole, the screening of $\omega_{\rm MB}^{(1,2)}$ is due to the source term, while that of $\omega_{\rm MB}^{(j\ge 3)}$ is due to poles of the Wronskian~\cite{Teukolsky:1972my,Teukolsky:1973ha}.  However, when using the Sasaki-Nakamura approach~\cite{Sasaki:1981sx,Sasaki:1981kj,Hughes:1999bq}, the entire screening arises from the Wronskian. 

\subsection{Dynamically excited direct waves}

The essential dynamical mechanism for direct-wave generation was identified more recently by Oshita {\it et al.}~\cite{Oshita:2025qmn}: although a stationary source oscillating exactly at Matsubara frequencies is screened by the potential barrier, a dynamically plunging source whose instantaneous frequency evolves toward a Matsubara frequency can still produce radiation that reaches infinity. Oshita {\it et al.}~\cite{Oshita:2025qmn} found numerical evidence for such a \textit{direct wave}. 

Similar to Refs.~\cite{Mino:2008at,Zimmerman:2011dx}, Oshita {\it et al.} interpreted the outgoing wave $\psi_4$ at infinity as the response of the black-hole spacetime to the driving of the plunging particle~\cite{Oshita:2025qmn}. In the frequency domain, we have
\begin{equation}
   \widetilde {\psi}_4(\omega) = {\frac{1}{i\omega B^{\rm in}(\omega)}}\cdot \widetilde{\mathcal{I}}(\omega) \equiv \widetilde{\mathcal{G}}_{\rm BH} \,\widetilde{\mathcal{I}}(\omega).
\end{equation}
Here, $\widetilde{\mathcal{I}}(\omega)$ is a trajectory integral involving the source and the horizon-ingoing homogeneous radial Teukolsky solution $R^{\rm in}$~\cite{Teukolsky:1973ha,SasakiReview}.
This integral contains factors of \mbox{$(\omega -\omega_{\rm MB}^{(1)})(\omega -\omega_{\rm MB}^{(2)})$}, as well as the asymptotics of the trajectory, which must rotate at $\Omega_H$ and accelerate toward the horizon with $\kappa_H$ in general relativity, as dictated by the property of the horizon;~\footnote{In Boyer--Lindquist coordinates, any regular timelike trajectory crossing the future horizon asymptotically follows the same coordinate trajectory as the ingoing principal null congruence.} this factor can be viewed as a driving force. 
The factor $\widetilde{\mathcal{G}}_{\rm BH} \equiv 1/(i\omega B^{\rm in})$ represents the response function of the black hole, where $B^{\rm in}(\omega)$ is an ingoing complex amplitude at past null infinity such that the ingoing amplitude at the future horizon is unity.
For fixed angular $(\ell,m)$ modes, it contains zeros at both the prograde and retrograde QNM frequencies~\cite{Leaver1985,BertiReview:2009}, $\omega_{\ell mn}$ and $\omega_{\ell m n }^{\rm R}$ ($n$ is the overtone number), and poles at the third and higher Matsubara frequencies~\cite{Mano:1996vt,SasakiReview,Motohashi2026,nakamoto2026}:
\begin{equation}
    B^{\rm in}(\omega) \sim \frac{\prod_{n} \left(\omega-\omega_{\ell m n }\right) \prod_{n} \left(\omega-\omega_{\ell m n }^{\rm R}\right)}{\prod_{j\geq 3} \left(\omega-\omega^{(j)}_{\rm MB}\right)} \, .
\end{equation}
In the time domain, we can write $\psi_4$ as a convolution of the time-domain response function and the time-domain source function: 
\begin{equation}
    \psi_4(t) =\int_{-\infty}^{+\infty} \mathcal{G}_{\rm BH}(t-t') \, \mathcal{I}(t') \, dt' .
\end{equation}
At earlier times, $\psi_4$ can be described as the sum of a QNM expansion (homogeneous part) and a driven response to $\mathcal{I}$ (inhomogeneous part).   
In the regime where $\widetilde{\mathcal{I}}(\omega)$ can be approximated by a rational function with a set of poles, $\widetilde{\psi}_4$ can be re-expanded as a sum of QNMs and source oscillations, subject to screening at the Matsubara frequencies due to the poles of $B^{\rm in}$. 

The direct-wave framework adopts a different perspective to probe the driving mechanism. 
To preserve the product (or driving-response) form of the waveform and extract information about the plunge trajectory --- particularly its final asymptotics --- Oshita {\it et al.} proposed removing the first few dominant QNMs using {\it rational filtering}~\cite{Ma:2022wpv,Ma:2023cwe,Ma:2023vvr}. This amounts to {\it multiplying} the frequency-domain waveform with a rational function of frequency, as further discussed in Sec.~\ref{subsec:rational}. Note that this differs from previous analyses, where the ringdown waveform is decomposed into an additive sum of components. Schematically, the QNM-filtered waveform is given by
\begin{equation}
\label{def:DW}
   \widetilde \Psi_{\rm DW}(\omega) = \mathcal{F}(\omega)\widetilde{\psi}_4(\omega) \sim \underbrace{\prod_{\ell mn\in S} \frac{\omega-\omega_{\ell mn } }{\omega-\omega_{\ell mn}^{*}}}_{\mathcal{F}(\omega)} \underbrace{\frac{\widetilde{\mathcal{I}}(\omega) }{i\omega B^{\rm in}(\omega)}}_{\widetilde{\psi}_4 (\omega)} \, ,
\end{equation}
where $\mathcal{F}$ is the rational filter, $S$ is the set of modes being filtered out, and $*$ denotes the complex conjugate. Having poles at complex conjugates of the zeros makes $|\mathcal{F}(\omega)|=1$ at all real frequencies, and therefore the rational filter introduces no new spectral power on the real frequency axis. 
It instead reorganizes information contained in the product $\widetilde{\mathcal{G}}_{\rm BH}(\omega)\widetilde{\mathcal{I}}(\omega)$.
The resulting $\Psi_{\rm DW}$ should therefore be interpreted as a source-sensitive representation of the waveform, rather than as an additional propagating component of the unfiltered signal~\cite{ma2026}.

For moderately high remnant spins (e.g., $\chi_f \gtrsim 0.5$), the Matsubara poles of $B^{\rm in}$ approximately cancel the (prograde) zeros associated with the third and higher QNM overtones (see Appendix~\ref{app:zeros_poles} and Refs.~\cite{Motohashi2026,weller2026}). It is therefore reasonable to apply a minimum set of rational filters that removes only the fundamental mode and the first two overtones ($n=0,1,2$). This approximately eliminates the frequency dependence of $B^{\rm in}$ and reveals features of $\widetilde{\mathcal{I}}(\omega)$, including the time-domain evolution of its complex frequency.  Note that the pole--zero cancellation becomes nearly exact for extremal black holes or at high overtones for moderate-to-high-spin black holes, giving rise to {\it pole skipping}~\cite{chen2010real,yang2013branching,yang2013quasinormal}.  Toward the other end, this cancellation fails completely in the Schwarzschild case, explaining the qualitatively different behavior observed by Ma and Wang~\cite{ma2026}. 

Filtering out only the $n=0,1,2$ modes naturally raises the question of whether the resulting signal is simply the $n=3$ QNM overtone. This is a legitimate concern that can be addressed through the subtle pole--zero structure of the Kerr response, as discussed in Appendix~\ref{app:zeros_poles}.

\begin{figure*}[t]
    \includegraphics[width=\textwidth]{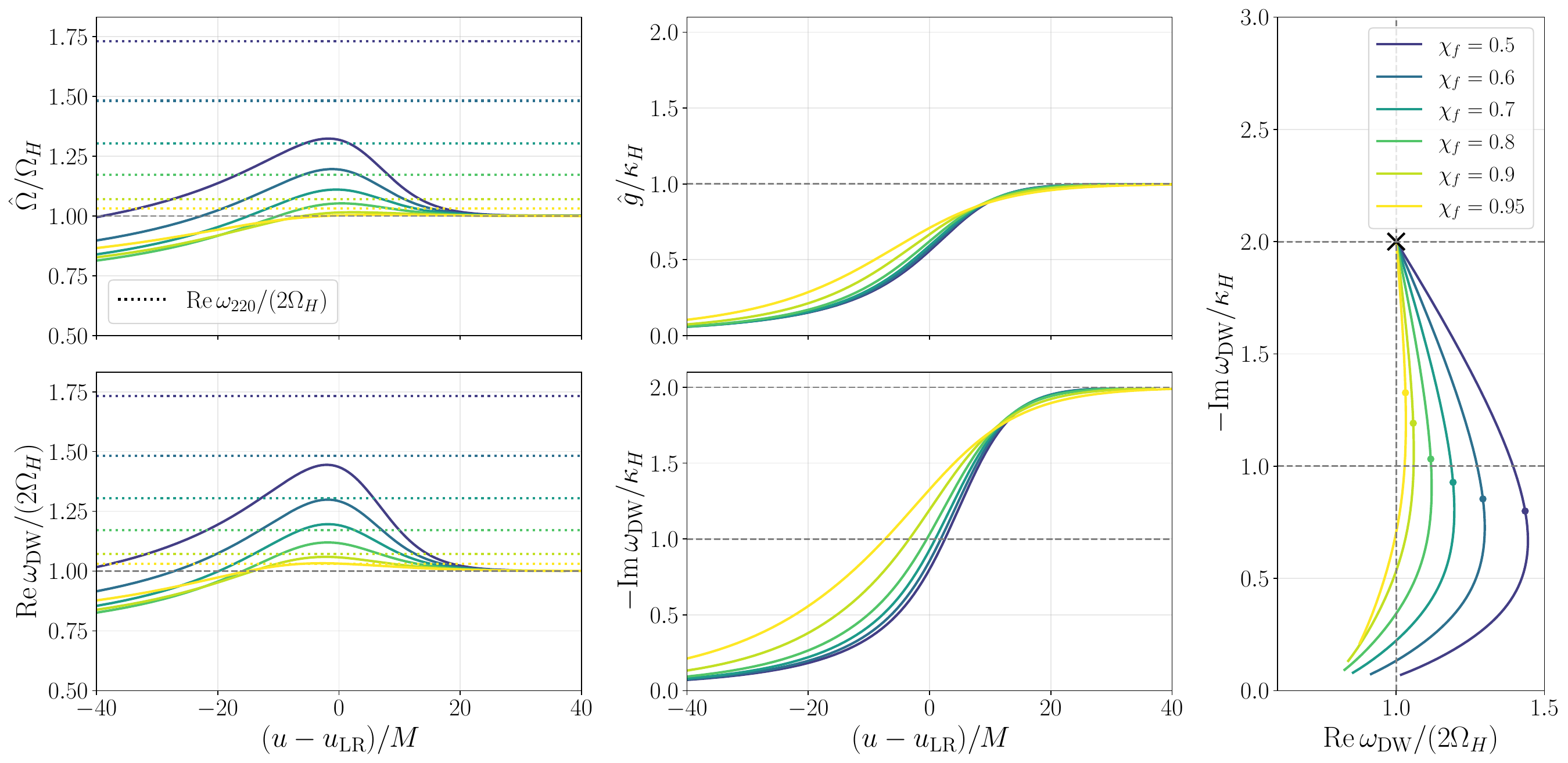}
    \caption{Analytical model predictions for the evolution of the direct-wave complex frequencies, unscreened $\omega_G$ [Eq.~\eqref{eq:omegaG}; upper] and screened $\omega_{\rm DW}$ [Eq.~\eqref{eq:omega_dw}; lower], emitted by particles plunging into Kerr black holes after an infinitesimal deviation from the equatorial ISCO, following Oshita {\it et al.}~\cite{Oshita:2025qmn}. Different colours correspond to the remnant spins indicated in the legend. The retarded time is shifted such that $u-u_{\rm LR}=0$ when the trajectory crosses the prograde equatorial light ring. 
    The upper-left panel shows the normalized unscreened frequency $\hat{\Omega}/\Omega_H$; the horizontal dotted lines indicate $\mathrm{Re}\,\omega_{220}/(2\Omega_H)$ for the corresponding Kerr spins. 
    The lower-left panel shows the resulting real part of the screened direct wave, $\mathrm{Re}\,\omega_{\rm DW}/(2\Omega_H)$. 
    For lower spins, $\mathrm{Re}\,\omega_{\rm DW}$ initially increases over time and substantially overshoots $2\Omega_H$ before returning asymptotically toward this value. With higher spins, the overshoot is much less pronounced. 
    The upper-middle panel shows the normalized unscreened decay scale $\hat{g}/\kappa_H$, while the lower-middle panel shows the instantaneous decay rate of the screened direct wave, $-\mathrm{Im}\,\omega_{\rm DW}/\kappa_H$. The latter increases toward the analytical model's asymptotic decay rate $2\kappa_H$, with the additional factor arising from the screening effect.     The right panel shows the same evolution parametrically in the complex-frequency plane. The dots mark the light-ring crossing, $u=u_{\rm LR}$, and the black cross marks the asymptote of Eq.~\eqref{eq:approx}, $\omega=2\Omega_H-2i\kappa_H$. 
    Note that the approximations underlying this analytical model do not capture the effective source's late-time asymptote, $m\Omega_H-3i\kappa_H$~\cite{weller2026}.
    The trajectories demonstrate that the finite-time direct-wave frequency need not equal its horizon asymptote: its real and imaginary parts evolve coherently toward that value during the plunge.}
    \label{fig:plunge}
\end{figure*}

\subsection{Analytic approximation for a plunging particle}

Using a steepest-descent analysis together with a near-horizon approximation of the Green's function, Oshita {\it et al.}~\cite{Oshita:2025qmn} proposed the analytic approximation for the inverse Fourier transform of $\widetilde\Psi_{\rm DW}$ in Eq.~\eqref{def:DW}, given by
\begin{equation}
    \Psi_{\rm DW}(u)
    \sim
    \bigl[\omega_G(u)-\omega_H\bigr]
    \exp\left[-i\int^u \omega_G(u')\,du'\right],
    \label{eq:approx}
\end{equation}
where
\begin{equation}
    \omega_H=m\Omega_H-i\kappa_H
\end{equation}
is the relevant horizon Matsubara frequency. Here, $u$ is the retarded time, traced back to the plunging trajectory through
$u(t)=t-r_*(t)$, if the orbit is parametrized by Boyer--Lindquist time $t$, and $r_*(t)$ is the Boyer--Lindquist tortoise coordinate of the particle at time $t$. 
The frequency $\omega_G$, defined at each point along the trajectory, is given by 
\begin{equation}
\label{eq:omegaG}
    \omega_G = 2\hat\Omega - i \hat g
    \,,\quad 
\hat\Omega = \frac{\beta\Omega_H + \Omega}{1+\beta}\,,\quad 
\hat g = \frac{2\beta\kappa_H}{1+\beta},
\end{equation}
where $\beta=-dr_*/dt$, and $\Omega = d\phi/dt$ with $\phi$ the azimuthal angle. More specifically, the exponential factor in Eq.~\eqref{eq:approx} arises from $\mathcal{I}$, while all other multiplicative factors involving QNM and Matsubara frequencies are assumed slowly varying and should be evaluated at $\omega = \omega_G(u)$. Since $\omega_G$ approaches $\omega_{\rm MB}^{(1)} =\omega_H$, the only significantly time-dependent factor is the screening factor of $[\omega_G(u) -\omega_H]$.~\footnote{Time dependence of factors like $\omega_G(u) - \omega_{\ell mn }^{*}$ are ignored since $\omega_G$ is far from $\omega_{\ell mn}^{*}$ on the complex plane.}

At this stage, it is important to distinguish three frequencies: (i) the relevant stationary horizon Matsubara frequency $\omega_H$ appearing in the screening prefactor, to which $\omega_G(u)$ asymptotes, (ii) the dynamically evolving complex frequency $\omega_G(u)$ inside the exponential, and (iii) the dynamically evolving composite instantaneous complex frequency $\omega_{\rm DW}(u)$ of the full direct wave. The latter two are not fixed; the instantaneous complex frequency $\omega_G(u)$ evolves asymptotically toward
\begin{equation}
    \omega_G(u)\rightarrow m\Omega_H-i\kappa_H,
    \label{eq:omega_G}
\end{equation}
with a decay of $e^{-\kappa_H u}$, and we obtain
\begin{equation}
    \omega_{\rm DW} (u) \equiv i\dot\Psi_{\rm DW}/\Psi_{\rm DW} \rightarrow 2\Omega_H - 2i \kappa_H\,,
    \label{eq:omega_dw}
\end{equation}
within this approximation.
The prefactor $\omega_G(u)-\omega_H$ reflects the screening of the stationary Matsubara component and introduces an additional suppression as the source approaches the horizon. The asymptotic limit of $\omega_{\rm DW}$ therefore does not imply the appearance of a stationary observable mode at that frequency; rather, it describes the limiting behavior of an evolving signal whose amplitude is suppressed simultaneously. 
Fundamentally, the fact that $\omega_{\rm DW}$ approaches to the Matsubara frequency reflects the extreme frame dragging of the horizon (in its real part), as well as the characteristic exponential redshift feature of the horizon (in its imaginary part).  Lu {\it et al.}~\cite{Lu2026} applied this framework to the QNM-removed binary black hole merger signal GW250114~\cite{GW250114,GW250114com} and identified evidence for a direct-wave component in the data collected by the two Advanced LIGO detectors~\cite{aLIGO}.

Fig.~\ref{fig:plunge} shows the real and imaginary parts of $\omega_G$ and $\omega_{\rm DW}$ obtained from Oshita {\it et al.} using the steepest-descent approximation and an orbit infinitesimally perturbed away from the innermost stable circular orbit (ISCO). The upper panels show $\hat{\Omega}/\Omega_H$ and $\hat{g}/\kappa_H$ as functions of retarded time $u$, while the lower panels show the corresponding real and imaginary parts of the full direct-wave frequency $\omega_{\rm DW}$.
In the physical picture developed by Oshita {\it et al.}~\cite{Oshita:2025qmn}, the direct wave initially follows the dynamics of the plunging source. As the source approaches the horizon, its angular frequency approaches the horizon rotation frequency. At the same time, screening near the Matsubara frequencies suppresses the outgoing wave more rapidly than a simple exponential with a constant decay rate, producing an increasing instantaneous damping rate. While the direct wave is tied to the dynamics of the plunging source, its late-time asymptotic behavior reflects the properties of the remnant horizon.

This analytical model described by Eq.~\eqref{eq:approx} captures the qualitative picture, but does not fully reproduce the effective source's late-time asymptotic.
The limitations arise from several simplifying assumptions, and its agreement with the numerical results demonstrated in Ref.~\cite{Oshita:2025qmn} is only qualitative. 
For example, the damping rate limit of $2\kappa_H$ in Eq.~\eqref{eq:omega_dw} is a consequence of such approximations. 
A more complete study of particle-plunge waveforms gives a nonvanishing leading coefficient, based on the near-horizon effective-source expansion, and the instantaneous complex frequency of the effective source approaches $m\Omega_H-3i\kappa_H$~\cite{weller2026}.
Indeed, when we study particle-plunge waveforms in Sec.~\ref{subsec:particle}, we find late-time discrepancies between the direct-wave evolution and the predictions of the simplified model in Eq.~\eqref{eq:approx}, along with its nominal asymptote in Eq.~\eqref{eq:omega_dw}; in particular, the damping rate instead approaches $3\kappa_H$. By this stage, however, the direct-wave amplitude has already become very small, and the approximations underlying Eq.~\eqref{eq:approx} no longer provide an accurate quantitative description. 

These limitations do not alter the main conclusions of the present analysis. The analytical model captures the qualitative physical picture that persists for comparable-mass binaries over the interval accessible in current NR waveforms. At later times, the numerical signal becomes dominated by noise; observationally, this final stage is unlikely to be measurable because of its strongly suppressed amplitude. The model therefore remains sufficient for qualitatively interpreting the reliably resolved NR evolution and assessing near-term observational prospects, although quantitative inference will ultimately require calibration against more complete NR predictions.

\subsection{Comparable-mass binaries}

\begin{figure}[t]
\centerline{
    \includegraphics[width=0.475\textwidth]{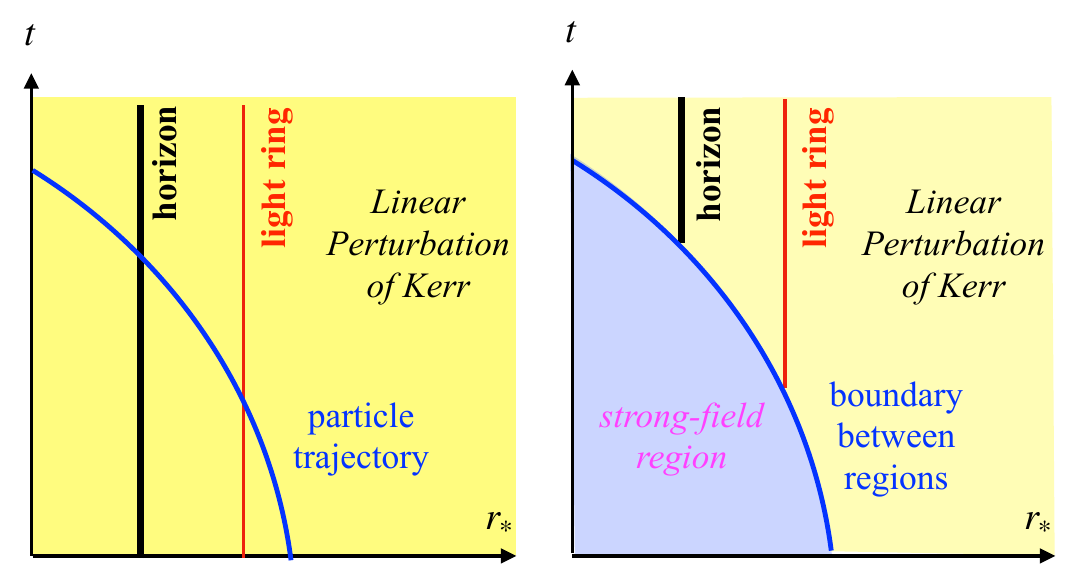}}
    \caption{Spacetime diagram comparing the plunge of a point particle in Kerr (left panel) and the merger of two black holes idealized as a nonspherical collapse (right panel). A boundary separating the strong- and weak-field perturbative regions can supply outgoing radiation to the weak-field region, mimicking the radiation from a plunging particle. }
    \label{fig:spacetime}
\end{figure}

The close connection between particle plunge and comparable-mass binary black hole merger has long been established through black-hole perturbation theory, numerical relativity, and effective-one-body descriptions~\cite{Damour:2007xr,Price:1994pm,Buonanno:1998gg,Buonanno:2000ef,Buonanno:2006ui,Berti:2007fi}. Nichols and Chen~\cite{Nichols:2011ih} developed a complementary hybrid picture in which the binary merger is modeled as an effective collapsing object whose nonspherical surface sources perturbations of the exterior black-hole spacetime, as illustrated in Fig.~\ref{fig:spacetime}. The argument of Zimmerman and Chen~\cite{Zimmerman:2011dx} also adopted this viewpoint. In this description, the direct wave encodes the oscillation and decay of curvature perturbations on the surface of the effective collapsing object as it approaches the horizon.

\section{Benchmark with NR and CCE waveforms}

In this section, we first review the approach for removing QNMs from NR waveforms using rational filters. We then discuss the properties of the extracted direct waves and compare them with particle-plunge examples and recent results in the literature.

\subsection{Rational filters and instantaneous frequency}
\label{subsec:rational}

We first apply rational filters~\cite{Ma:2022wpv,Ma:2023cwe,Ma:2023vvr} to remove QNMs from the waveform in the Fourier domain [see Eq.~\eqref{def:DW}]. For the $(2,2)$ spherical-harmonic component $h_{22}$,
we remove both prograde and retrograde QNMs contributing to the $(2,2)$ component.
The filtered waveform is written as
\begin{equation}
    \widetilde{z}(\omega)
    = \mathcal{F(\omega)}\widetilde{h}_{22}(\omega).
\end{equation}
Each factor in the rational filter is an all-pass filter on the real-frequency axis, preserving the Fourier-domain amplitude while modifying the phase to remove the targeted QNM contribution from the time-domain waveform. 

After inverse Fourier transforming the filtered waveform, we compute its instantaneous complex frequency directly as
\begin{equation}
    \omega(t)=i{\dot{z}(t)}/{z(t)}.
\end{equation}
Writing $z(t)=A(t)e^{-i\phi(t)}$, this gives
\begin{equation}
    \omega(t)=\dot{\phi}(t)+i{\dot{A}(t)}/{A(t)},
\end{equation}
such that $\mathrm{Re}\,\omega=\dot{\phi}$ is the instantaneous oscillation frequency and $-\mathrm{Im}\,\omega=-\dot{A}/A$ is the instantaneous decay rate. This construction therefore allows us to follow the continuous evolution of both the oscillation frequency and damping of the non-QNM residual, rather than characterizing it by a single complex frequency fitted over a finite-time interval.

In practice, we extract the instantaneous frequencies from the filtered Bondi news~\cite{Bondi:1962px,Sachs:1962wk} for the NR/CCE waveforms, to suppress constant offsets, secular drifts, and very-low-frequency artifacts; we use the filtered strain for particle-plunge examples.

\subsection{NR and CCE results}
\label{subsec:nr_cce}

\begin{figure*}
\includegraphics[width=0.9\textwidth]{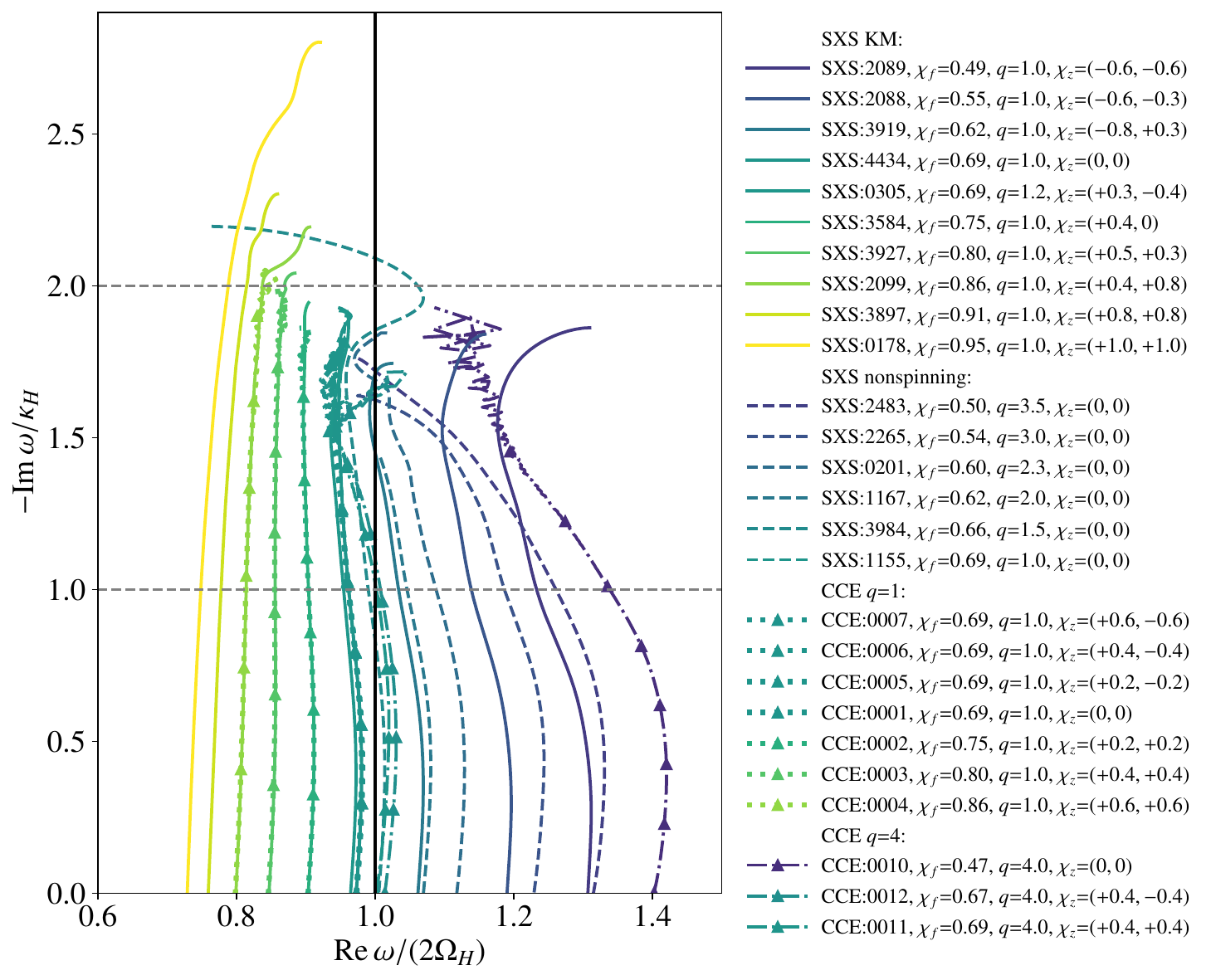}
\caption{Evolutions of the real and imaginary parts of the instantaneous direct-wave complex frequency from the filtered news function. Solid curves show the SXS binaries studied by Kankani and McWilliams in Ref.~\cite{kankani2026} (labelled ``SXS KM'' in the legend); dashed curves show SXS binaries with non-spinning progenitor black holes (``SXS nonspinning''); dotted and dash-dotted curves with triangle markers show CCE binaries with mass ratios $q=1$ and $q=4$, respectively. The entries in each $\chi_z$ array give the dimensionless spin components of the two progenitor black holes along the $z$ axis. The colour scale, from purple to yellow, indicates increasing final black-hole spin $\chi_f$. Each trajectory starts at the peak of the filtered news function and is truncated at the first local maximum of $|\mathrm{Im}(\omega_{\rm DW})|$. The endpoints mark the loss of numerical reliability and should not be interpreted as measurements of the asymptotic complex frequencies.}
\label{fig:all_clim_tracks}
\end{figure*}

For NR waveforms from the SXS catalog (obtained via extrapolation toward infinity, referred to as NR waveforms)~\cite{Boyle:2019kee,Scheel:2025jct,Boyle:2009vi}, we carry out two complementary studies, followed by a comparison with waveforms obtained using CCE (referred to as CCE waveforms). In this section, we adopt a rational filter baseline configuration to remove modes with angular multipole index $\ell\leq6$ and up to the second overtone ($n=0,1,2$) from the $(2,2)$ component, including both prograde and retrograde modes (see further discussions in Appendix~\ref{app:zeros_poles}). 

First, we analyze the ten systems considered in Ref.~\cite{kankani2026}. The detailed time evolution of the direct-wave amplitude, together with the evolution of the real and imaginary parts of the instantaneous direct-wave frequency, $\omega_{\rm DW}$, is provided in Appendix~\ref{app:detailed_tracks}. 
Figs.~\ref{fig:SXS_KM_1} and \ref{fig:SXS_KM_2} show that the extracted direct wave undergoes a clear transition to a different evolution pattern at a {\it transition time}, after which it no longer exhibits the evolution associated with the direct-wave component (shaded regions).
The middle panels show that, although the real part of the extracted direct-wave frequency, $\mathrm{Re}\,\omega_{\rm DW}$, generally differs from $2\Omega_H$ at earlier times, it evolves systematically toward this value. In particular, for remnant spins $\chi_f\lesssim 0.7$, $\mathrm{Re}\,\omega_{\rm DW}$ approaches $2\Omega_H$ from above (reaching as high as $\sim1.4\times2\Omega_H$), whereas for $\chi_f\gtrsim 0.7$, it approaches from below (down to $\sim 0.7\times2\Omega_H$). 
The right panels show that, after the peak of the direct-wave amplitude, the imaginary part of the extracted direct-wave frequency, $-\mathrm{Im}\,\omega_{\rm DW}$, evolves to values typically of order $\sim2\kappa_H$, with some reaching $\sim3\kappa_H$ for high-$\chi_f$ cases. 
Beyond the transition time, however, both the real and imaginary parts of the extracted $\omega_{\rm DW}$ cease to have a clear physical interpretation. We empirically define the transition time as the moment when the numerically extracted $|\mathrm{Im}\,\omega_{\rm DW}|$ reaches its first local maximum. It is indicated by a vertical line in the figures in Appendix~\ref{app:detailed_tracks}, with the post-transition region shaded. 
Here, we show the corresponding parametric trajectories in the $(\mathrm{Re}\,\omega_{\rm DW}, \mathrm{Im}\,\omega_{\rm DW})$ plane as solid curves in Fig.~\ref{fig:all_clim_tracks}.
The trajectories in Fig.~\ref{fig:all_clim_tracks} are terminated at the transition time. 

Second, we consider a set of nonspinning binaries with mass ratios increasing from $q=1$ to $q=3.5$, corresponding to remnant spins decreasing from approximately $\chi_f=0.68$ to $\chi_f=0.50$. Their parametric trajectories are shown as dashed curves in Fig.~\ref{fig:all_clim_tracks}, with the detailed numerical results presented in Fig.~\ref{fig:SXS_nonspinning}. In this set, the direct wave evolution trend is particularly clear, with the extracted complex frequency approaching the analytical limit in Eq.~\eqref{eq:omega_dw}. 
We caution, however, since the reliable trajectory terminates once the extraction loses numerical accuracy, these terminal values should not be interpreted as asymptotic decay rates or as evidence for a physical limit.
The lower terminal decay rates seen in some cases may therefore depend partly on when the direct-wave signal becomes indistinguishable from numerical noise.

Finally, we analyze ten CCE waveforms, which reproduce the same spin-dependent evolution seen in the extrapolated NR waveforms. Their trajectories are shown as dotted ($q=1$) and dash-dotted ($q=4$) curves with triangle marks in Fig.~\ref{fig:all_clim_tracks}; the corresponding numerical results are presented in Figs.~\ref{fig:SXS_CCE_1}--\ref{fig:SXS_CCE_2}. 
At sufficiently late times, however, high-frequency numerical noise becomes prominent as the direct-wave amplitude decays, limiting the reliable reconstruction of the instantaneous complex frequency.

Fig.~\ref{fig:all_clim_tracks} summarizes the complex-frequency evolution across all NR and CCE waveforms considered in this section. Despite differences in mass ratio, progenitor spins, and waveform-extraction method, the trajectories exhibit a common pattern: the real frequency evolves toward $2\Omega_H$ from above or below, depending on the remnant spin, while the instantaneous decay rate increases, reaching $2\kappa_H$ and sometimes $\sim3\kappa_H$. Each trajectory is truncated at the transition time, beyond which the decreasing direct-wave amplitude and increasing numerical contamination render the reconstructed instantaneous complex frequency unreliable.

This behavior can be understood qualitatively as a transition from the orbital motion inherited from the merger to horizon-dominated frame dragging. 
For remnants with $\chi_f\sim0.7$, characteristic of equal-mass, nonspinning binaries, the orbital frequency near merger is already comparable to $\Omega_H$ and transitions smoothly to the remnant horizon angular velocity~\cite{Buonanno:2006ui,Berti:2007fi}. The early-time direct-wave frequency therefore lies close to $2\Omega_H$, without requiring substantial subsequent evolution. For lower-spin remnants, the merger frequency exceeds $\Omega_H$, and the direct-wave frequency approaches $2\Omega_H$ from above as frame dragging becomes increasingly dominant. For higher-spin remnants, the merger frequency lies below $\Omega_H$, and the stronger frame dragging causes the direct-wave frequency to increase toward $2\Omega_H$ from below. 
In all cases, the trajectories record the transition from merger dynamics and evolve toward the same horizon-controlled asymptotic behavior.

We now discuss several irregularities in Fig.~\ref{fig:all_clim_tracks}. In SXS:2089 and SXS:2088, $\mathrm{Re}\,\omega_{\rm DW}$ initially decreases toward $2\Omega_H$ before briefly turning upward. These turning features correspond to the bumps in the two upper middle panels of Fig.~\ref{fig:SXS_KM_1} and are likely to arise from spurious oscillations arising near the end of the interval over which the direct wave can be reliably extracted. Similarly, for SXS:3984, the middle panel in the second row of Fig.~\ref{fig:SXS_nonspinning} shows that the transition time occurs when the spurious component emerges. 
Finally, for the $\chi_f>0.8$ cases, we see that $\mathrm{Re}\,\omega_{\rm DW}$ increases toward $2\Omega_H$ but does not reach it before the transition time. 
Beyond this point, it appears to continue approaching $2\Omega_H$, although the evolution becomes increasingly oscillatory and noisy, as shown in the middle panels of Fig.~\ref{fig:SXS_KM_2}.

Within the accuracy of our direct-wave extraction in NR and CCE waveforms, the trends in Fig.~\ref{fig:all_clim_tracks} are qualitatively consistent with the analytic approximation of Eq.~\eqref{eq:approx} for trajectories plunging from the ISCO, shown in the right panel of Fig.~\ref{fig:plunge}. 
For low $\chi_f$, $\mathrm{Re}\,\omega_{\rm DW}$ initially rises well above $2\Omega_H$ before turning back toward it, while $-\mathrm{Im} \, \omega_{\rm DW}$ generally increases with time. 
For high $\chi_f$, however, the numerical results show that $\mathrm{Re}\,\omega_{\rm DW}$ remains below $2\Omega_H$ and gradually approaches it, whereas $-\mathrm{Im}\,\omega_{\rm DW}$ tends to increase toward $3\kappa_H$.
We emphasize that the analytic approximation in Eq.~\eqref{eq:approx} provides only a simplified model of the direct wave. Quantitative differences from the numerical results are therefore expected and do not affect the qualitative physical interpretation of the extracted direct-wave signal.

As illustrated in Fig.~\ref{fig:spacetime}, the direct wave initially tracks perturbations sourced by the collapsing boundary between the strong-field and perturbative regions. 
Its complex frequency evolves toward horizon-controlled behavior, with the real part tending toward $2\Omega_H$ and the decay rate increasing over the numerically resolved time interval.
In this picture, the direct wave traces the formation of the black-hole horizon, but is hidden beneath the dominant QNM emission associated with the light ring.

\subsection{Particle plunges from ISCO}
\label{subsec:particle}

\begin{figure}
    \includegraphics[width=0.5\textwidth]{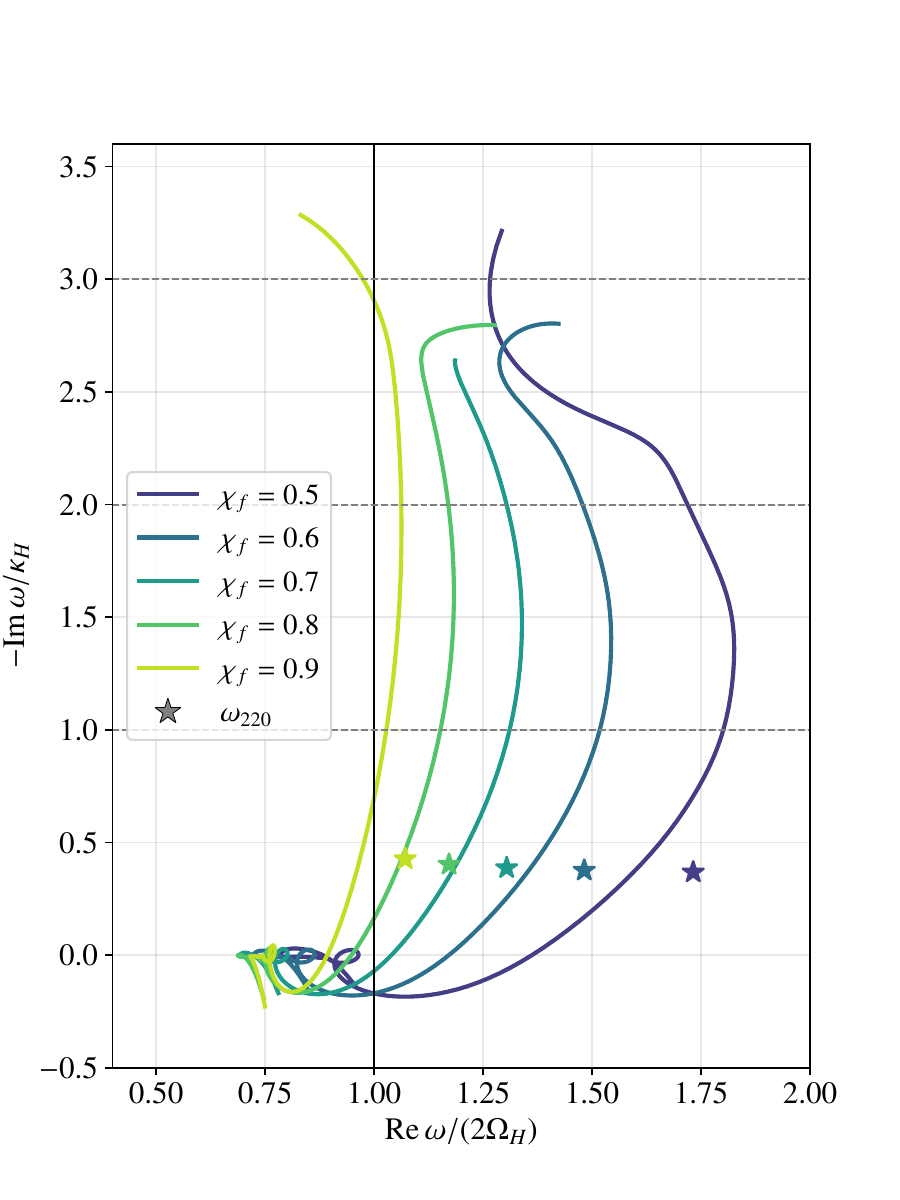}
    \caption{Evolution tracks of the real and imaginary parts of the filtered direct-wave complex frequency for ISCO plunges into black holes with spins $\chi_f$ ranging from 0.5 to 0.9. We retain only the portions of the tracks satisfying the empirically chosen adiabaticity criterion $\left|d\omega/du\right|\leq 0.15\,|\omega|^2$.
    }
\label{fig:emri:climb}
\end{figure}

As noted by Oshita {\it et al.}~\cite{Oshita:2025qmn}, the analytical approximation is intended to provide a qualitative description of the filtered direct wave. Here, we perform a small-scale parameter study in which we extract the instantaneous complex-frequency evolution directly from filtered ISCO-orbit waveforms and compare it with the analytical approximation. 
We consider final spins $\chi_f=0.5,\,0.6,\,0.7,\,0.8,$ and $0.9$, and apply rational filters to remove the $\ell=2$ prograde and retrograde QNMs with overtone numbers $n = 0,1,2$ contributing to the $(2,2)$ component. From the filtered strain $\mathcal{F}h$, we reconstruct the evolution of the real and imaginary parts of the direct-wave frequency. 
Fig.~\ref{fig:emri:climb} shows the resulting trajectories in the ($\mathrm{Re}\,\omega_{\rm DW}$,\,$\mathrm{Im}\,\omega_{\rm DW}$) plane.
The extracted direct-wave complex frequency depends on the set of modes removed by the rational filter (see more details in Appendix~\ref{app:zeros_poles}); a systematic study of this dependence will be presented in future work. Here, we note that removing modes up to the second overtone yields a frequency evolution qualitatively similar to that found in the above numerical waveforms.  

Note that the ISCO trajectory considered here, with an infinitesimal initial plunge, naturally yields higher values of $\mathrm{Re}\,\omega_{\rm DW}$ at fixed $\mathrm{Im}\,\omega_{\rm DW}$ than a plunge with a finite infall velocity at the ISCO, the latter providing a more appropriate comparison with comparable-mass mergers. This helps explain the early-time difference between Figs.~\ref{fig:all_clim_tracks} and \ref{fig:emri:climb}.
At late times, corresponding to larger decay rates, we also find that $-\mathrm{Im}\,\omega_{\rm DW}$ approaches $3\kappa_H$, exceeding the $2\kappa_H$ limit of the analytical model [cf. Eqs.~\eqref{eq:approx}--\eqref{eq:omega_dw}], reflecting the limitations of the underlying approximations in the regime where the signal has become very weak. This also explains the late-time difference between Figs.~\ref{fig:plunge} and \ref{fig:emri:climb}.

\subsection{Summary and comparison with recent literature}

Our studies in the preceding sections have three conceptual implications. (i) The rational-filter analysis does not directly {\it reach} the horizon. (ii) Nevertheless, the QNM-filtered direct-wave component captures the particle's approach to the horizon: the turning behavior of $\mathrm{Re}\,\omega_{\rm DW}$ reflects frame dragging, while the increase in $-\mathrm{Im}\,\omega_{\rm DW}$ reflects gravitational redshift and screening. (iii) Quantitative inference of horizon properties requires calibration against waveforms predicted by general relativity. Once such a calibration is established, the direct wave can probe aspects of binary black hole merger and remnant formation that are complementary to those encoded in the QNMs.

The instantaneous complex-frequency evolution obtained in the previous subsection is consistent with the effective frequencies inferred in Ref.~\cite{kankani2026} over fixed time windows near the waveform peak, which sample a particular stage of this evolution.
In particular, as $\chi_f$ increases from 0.5 to 0.95, the real part reported in Ref.~\cite{kankani2026} ranges from approximately $1.4\times2\Omega_H$ to $0.75\times2\Omega_H$. These values agree with our instantaneous frequencies at earlier times, while our analysis further reveals their subsequent evolution toward $2\Omega_H$. Similarly, the imaginary part reported in Ref.~\cite{kankani2026} ranges from approximately $0.75\kappa_H$ to $1.1\kappa_H$ for symmetric time windows around the waveform peak, but generally increases when the windows are extended to later times, consistent with the evolution found in the previous subsection.

Ref.~\cite{kankani2026} also applied the analytic approximation to SXS:0178 and found that it does not reproduce the extracted direct-wave frequency, with real part remaining below $2\Omega_H$. This case is shown in the bottom panel of Fig.~\ref{fig:SXS_KM_2}. Our analysis indicates that $\mathrm{Re}\,\omega_{\rm DW}$ continues to evolve toward $2\Omega_H$, while $-\mathrm{Im}\,\omega_{\rm DW}$ approaches $\sim3\kappa_H$. Thus, we interpret the disagreement reported in Ref.~\cite{kankani2026} as the limitations of the simplified model, rather than as evidence against the physical interpretation of the direct wave itself.

Our results complement the recent analyses presented in Refs.~\cite{dyer2026} and \cite{han2026}. Using CCE waveforms, Ref.~\cite{dyer2026} found that including an {\it additive} direct-wave component improves the modeling of early-time ringdown, while free-frequency fits drift by approximately $10\%$--$20\%$ as the start time is varied rather than remaining fixed at $m\Omega_H$. This behavior is consistent with our finding that the finite-time instantaneous frequencies offset from the horizon frequency, and it arises naturally from the continuous, spin-dependent evolution of the direct-wave frequency. Ref.~\cite{han2026} independently showed that the observed damping is also dynamical.
They demonstrated that the combined effects of screening, convolution with a redshift-stretched plunge source, and finite-window fitting map the local horizon redshift onto an effective envelope damping rate, $\gamma_{\rm eff}$, and obtained $\gamma_{\rm eff}/\kappa_H\simeq0.6$ for GW250114. 
Although their finite-window envelope slope is not identical to the local instantaneous decay rate studied here, both analyses demonstrate that a damping rate inferred over a finite time window should not be interpreted directly as the underlying horizon properties. As proposed in Ref.~\cite{han2026}, 
quantitative inference of horizon properties from the direct wave therefore requires calibrating the mapping between the measured frequency and damping evolution against NR waveforms. Such calibration also underlies measurements of remnant properties using direct waves, such as those performed in Ref.~\cite{chung2026}.

\section{Conclusion}

In this work, we have clarified the physical interpretation of the direct wave and its connection to the remnant black-hole horizon. Because the event horizon is the boundary of the causal past of future null infinity, no signal emitted from the horizon itself can reach a distant observer. Any ``direct'' information about it must instead be inferred from radiation generated as matter or spacetime perturbations approach it. The direct wave, understood here as a source-sensitive, response-filtered representation of the merger waveform that exposes information about the plunge and remnant formation, serves as such an asymptotic messenger. 
Indeed, the complex-frequency evolution of direct waves from NR closely resembles that of radiation from a particle plunging into a Kerr black hole.

Rather than behaving as a stationary damped sinusoid with a fixed complex frequency, the direct wave is characterized by an evolving complex frequency, $\omega_{\rm DW}(t)$, that tracks the plunge dynamics before approaching the horizon-controlled asymptote at sufficiently late times. This finite-time evolution is an intrinsic feature of the direct-wave framework already developed by Oshita {\it et al.}~\cite{Oshita:2025qmn} and applied by Lu {\it et al.}~\cite{Lu2026}.

Using QNM-removed NR and CCE waveforms, we reconstruct the evolution of the instantaneous complex frequency of the direct wave across a range of binary systems. The resulting trajectories depend systematically on the remnant spin: for lower spins, the real part approaches $2\Omega_H$ from above, whereas for higher spins, it approaches from below, consistent with the expected influence of frame dragging. 
Meanwhile, the instantaneous decay rate generally increases, reaching $\sim 2\kappa_H$, and approaching $3\kappa_H$ in the best-resolved cases, before losing numerical accuracy. 
In the approximation that the filtered direct wave follows the effective source trajectory in the complex-frequency plane, and when the direct-wave component remains sufficiently well resolved, the direct-wave frequency approaches the theoretical asymptote
\begin{equation}
\omega_{\rm DW}\rightarrow 2\Omega_H-3i\kappa_H.
\end{equation}
At later times, contamination from numerical noise and residual waveform components renders the extracted instantaneous complex frequency unreliable. These results lead to the following main conclusions:
\begin{enumerate}
    \item 
    The direct wave evolves dynamically; its evolution reflects the continuing collapse of the particle or field perturbations into the remnant black hole. Although current observations and numerical waveforms probe only part of this evolution, its late-time behavior reflects the final stages of collapse and hence the formation of the horizon: the real part of its instantaneous frequency approaches $2\Omega_H$, while its instantaneous decay rate increases with time.
    \item For non-precessing, comparable-mass binaries with $\chi_f\sim0.7$, the early-time direct-wave frequency is already close to $2\Omega_H$, because the binary orbital frequency  transitions smoothly to the angular velocity of the remnant horizon. The agreement found by Lu {\it et al.}~\cite{Lu2026} is therefore not coincidental: for this special class of binaries, general relativity predicts that an early-time measurement of the direct-wave frequency can provide a well-motivated probe of the horizon frequency, even when the signal-to-noise ratio is insufficient to track the subsequent evolution. For remnants with other spins, inferring horizon properties requires modeling the complete frequency trajectory, including its late-time approach to the horizon-controlled asymptote, and using NR simulations to calibrate the mapping between this asymptote and the observationally accessible earlier-time evolution. In particular, the $\chi_f$-dependent evolution of $\mathrm{Re}\,\omega_{\rm DW}$ provides a signature of near-horizon frame dragging, while the increase in $-\mathrm{Im}\,\omega_{\rm DW}$ reflects gravitational redshift and screening.
    \item Rational filtering must be applied with caution. Filtering additional QNM overtones suppresses the waveform near the corresponding Matsubara frequencies, thereby increasing the relative importance of numerical noise and other waveform artifacts. 
    Following the product structure $\widetilde \psi_4(\omega)= \widetilde{\mathcal{G}}_{\rm BH} \,\widetilde{\mathcal{I}}(\omega)$ with $\widetilde{\mathcal{G}}_{\rm BH}={1}/{(i\omega B^{\rm in})}$, our aim is to reveal the dynamical features of the source $\widetilde{\mathcal{I}}(\omega)$ by separating it from the black-hole response $\widetilde{\mathcal{G}}_{\rm BH}(\omega)$. As illustrated in Fig.~\ref{fig:all_clim_tracks}, filtering the fundamental mode and the first two overtones ($n=0,1,2$) already largely compensates for the frequency-dependent phase of $1/B^{\rm in}(\omega)$ along the real-frequency axis. Filtering additional overtones therefore provides limited benefit while increasing the suppression near the corresponding Matsubara frequencies.
    \item The pole--zero structure of the black hole response function also provides a complementary interpretation of the direct wave's asymptotic behaviour. Near-horizon source dynamics need not appear as a separate oscillation at the driving frequency: when a source pole coincides with a zero of the relevant response factor, instead of being fully screened, the driven contribution can be {\it reorganized} into neighboring QNM poles due to the effect of antiresonance driving~\cite{weller2026}. The appearance of a QNM-like response therefore does not, by itself, imply that the late plunge has ceased to influence the waveform. In general, source dynamics that deviate from near-horizon asymptotics predicted by general relativity will produce frequency components associated with those dynamics, which are distinct from both QNM and Matsubara frequencies; for moderate-to-high spins, these components can dominate over the QNM contributions. In this sense, the direct-wave component can be used not only to probe near-horizon source dynamics, but also provide additional support for interpreting the QNM-like wave in general relativity as being driven by the plunging source. 
\end{enumerate}

Our analysis uses NR waveforms generated by the evolution of a dynamical black-hole spacetime.
Once the remnant is sufficiently close to Kerr, its perturbations can, in principle, be propagated into the bulk of the perturbed Kerr spacetime and related to the underlying spacetime geometry~\cite{ma2022gravitational}. The direct wave therefore offers a route toward connecting an observable waveform component to the dynamics of black-hole and horizon formation.

The present study has several limitations. We have not systematically characterized direct-wave evolution across the full binary parameter space, including its dependence on mass ratio, progenitor spins, precession, higher-order modes, waveform-extraction methods, and numerical resolution. Because the approach to the horizon-controlled asymptote is part of the nonlinear formation of the remnant, quantitative inference of horizon properties will ultimately require calibration against NR simulations and comparison with general-relativistic waveform predictions. This requirement is not unique to the direct-wave framework, but is common to gravitational-wave inference more broadly. Future work should develop a complete model of the evolution, from its early-time behavior through the transition to the late-time asymptote, while accounting for both binary parameters and waveform systematics.

\begin{acknowledgments}
We gratefully acknowledge the Simulating eXtreme Spacetimes (SXS) Collaboration for making their numerical-relativity waveforms publicly available, and the SpECTRE Collaboration for providing the CCE waveforms.
P.C., A.L., and Y.C.\ are supported by the Brinson Foundation, the Simons Foundation (Award Number 568762), and by US NSF Grants PHY-2309211 and PHY-2309231.
The authors acknowledge the support by the Australian Research Council (ARC) Centre of Excellence for Gravitational Wave Discovery (OzGrav), Project Number CE230100016. L.S. is also supported by the ARC Discovery Early Career Researcher Award, Project Number DE240100206. 
Research at Perimeter Institute is supported in part by the Government of Canada through the Department of Innovation, Science and Economic Development and by the Province of Ontario through the Ministry of Colleges and Universities. A.L. is grateful for support from the Fannie and John Hertz Foundation in the form of a Hertz Fellowship.
\end{acknowledgments}

\appendix

\section{Response pole--zero structure and rational filters}
\label{app:zeros_poles}

\renewcommand{\thefigure}{A\arabic{figure}}
\setcounter{figure}{0}

One ambiguity in the definition of the direct wave $\widetilde\Psi_{\rm DW}$ concerns how many QNMs should be removed using the rational filter $\mathcal{F}$. Applying additional filters associated with higher overtones can affect the extraction of the direct-wave frequency evolution, even though fewer QNMs are present in the relevant waveform interval, because these filters introduce further suppression near the corresponding Matsubara frequencies.  This issue is related to the finding in Ref.~\cite{ma2026}: in the Schwarzschild case, direct waves extracted after removing a large number of overtones do not directly exhibit the expected horizon-frequency behavior, even though the plunge trajectory, which has the appropriate asymptotic near-horizon behavior, can still be recovered. This highlights the need for a better understanding of the black hole Green's function.

\subsection{Motivation for the choice of rational filters}

In this work, we have empirically found that removing only the QNMs with $n=0,1,2$, for both NR and particle plunging waveforms, yields behavior more consistent with the analytical model of Eq.~\eqref{eq:approx}. Here, we justify this prescription using the analytical pole--zero structure of $B^{\rm in}$. First, we recognize that the inverse Fourier transform of $\widetilde{\mathcal{I}}(\omega)$ in Eq.~\eqref{def:DW} contains contributions whose complex frequencies asymptote to the horizon frequencies, subject to screening at the first two Matsubara frequencies. Meanwhile, $B^{\rm in}$ has zeros at the QNM frequencies and poles at the third and higher Matsubara frequencies, $j\geq3$.

\begin{figure*}
    \includegraphics[width=\textwidth]{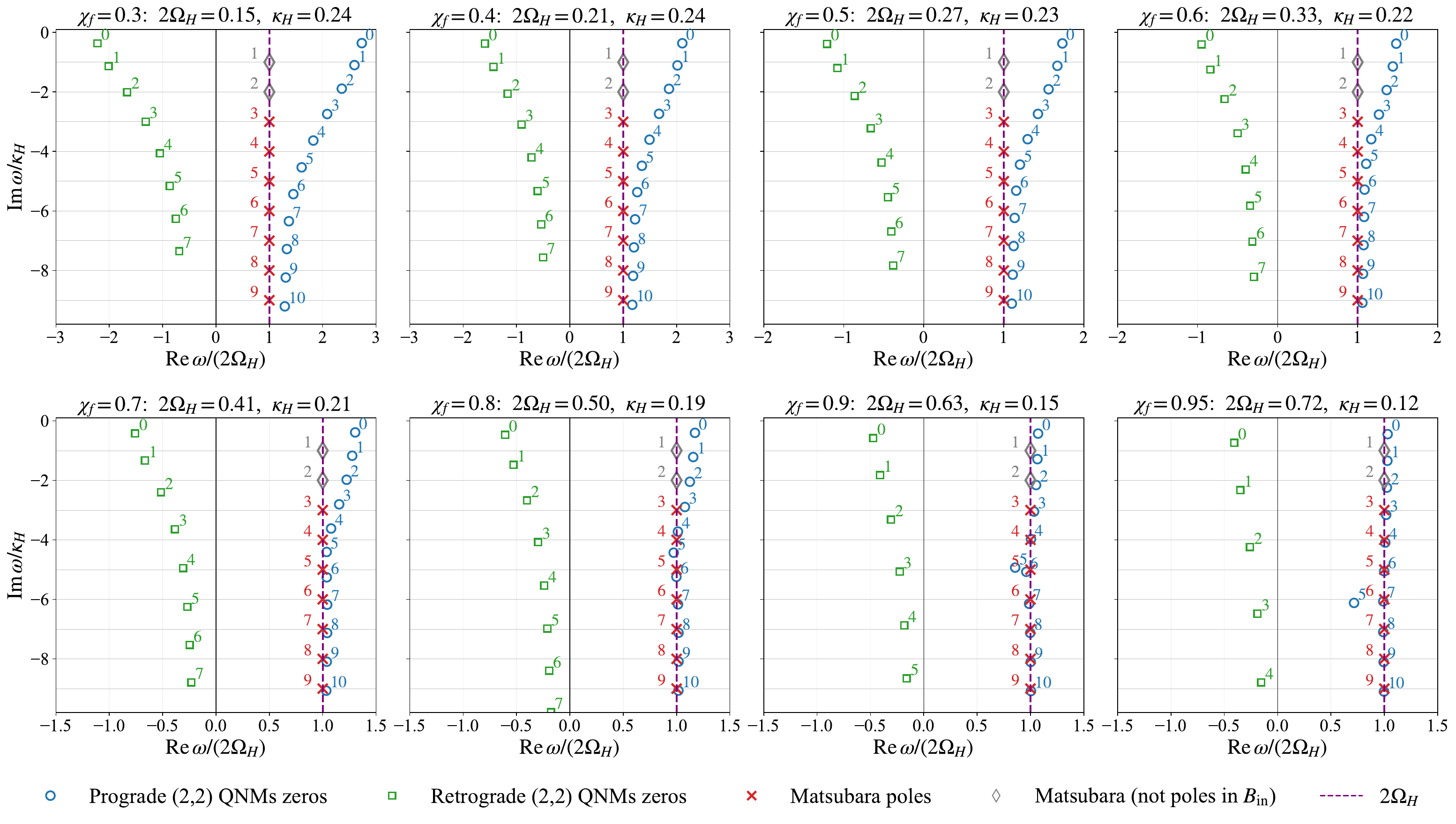}
    \caption{Zeros and poles of $B^{\rm in}(\omega)$ in the lower half of the complex-frequency plane for various values of $\chi_f$ ($2\Omega_H$ and $\kappa_H$ are expressed in units of the inverse black-hole mass).  Zeros of $B^{\rm in}(\omega)$ correspond to prograde (blue circles) and retrograde (green squares) QNMs, indexed by overtone $n$, while the poles correspond to Matsubara frequencies with $j\geq3$ (red crosses). The $j=1,2$ Matsubara frequencies, shown as gray diamonds, are not poles of $B^{\rm in}(\omega)$. For larger values of $\chi_f$, the Matsubara poles and QNM zeros approach one another, and their effects approximately cancel along the real-frequency axis when the differences between their real parts are much smaller than the magnitudes of their imaginary parts. The special $n=5$ overtone is an exception. The pole-zero pairing phenomenon motivates the use of rational filters that remove only the overtones with $0\leq n\leq2$. \label{fig:zp}} 
\end{figure*}

In Fig.~\ref{fig:zp}, we show the prograde and retrograde QNM frequencies as blue circles and green squares, respectively, the first two Matsubara frequencies as gray diamonds, and the third and higher Matsubara frequencies as red crosses. For larger spins, e.g., $\chi_f\gtrsim0.5$, the QNM zeros and the Matsubara poles with $j\geq3$ approximately coincide, with differences in their real parts much smaller than the average magnitudes of their imaginary parts. Consequently, along the real-frequency axis, the frequency dependence of $B^{\rm in}$ is effectively governed only by the $n=0,1,2$ QNM zeros. An exception occurs for $\chi_f>0.9$, where the $n=5$ QNM has an isolated feature at a relatively high decay rate. The asymptotic correspondence between higher-order prograde QNM overtones and Matsubara frequencies was studied in detail for extremal black holes in Refs.~\cite{Keshet:2007be,Detweiler:1980gk,Hod:2008zz,yang2013branching,yang2013quasinormal}; it is somewhat surprising that this correspondence remains approximately valid even at moderate spins. This structure explains why it is sufficient to filter out only the $n=0,1,2$ QNMs: the screening associated with the $j\geq3$ Matsubara poles is effectively canceled, or ``protected,'' by the corresponding QNM zeros in $B^{\rm in}$.

The approximate cancellation illustrated in Fig.~\ref{fig:zp} can be understood from a local representation of the black-hole response.
Near a QNM frequency $\omega_{\ell mn}$ and a neighboring Matsubara frequency $\omega_{\rm MB}^{(j)}$, with $j\geq 3$, the relevant factor in $\widetilde{\mathcal{G}}_{\rm BH}=1/(i\omega B^{\rm in})$ takes the schematic form
\begin{equation}
    \widetilde{\mathcal{G}}_{\rm BH}^{\rm loc}(\omega)
    =
    A(\omega)
    \frac{\omega-\omega_{\rm MB}^{(j)}}
         {\omega-\omega_{\ell mn}},
    \label{eq:local_pole_zero}
\end{equation}
where $A(\omega)$ varies slowly in the neighborhood of the pole--zero pair.
The residue of the QNM pole is therefore
\begin{equation}
    \mathop{\rm Res}_{\omega=\omega_{\ell mn}}
    \widetilde{\mathcal{G}}_{\rm BH}^{\rm loc}
    =
    A(\omega_{\ell mn})
    \left(\omega_{\ell mn}-\omega_{\rm MB}^{(j)}\right).
    \label{eq:pole_zero_residue}
\end{equation}
Thus, when $\omega_{\ell mn}\simeq\omega_{\rm MB}^{(j)}$, the QNM remains part of the spectrum, but its independent resonant imprint in this response factor is suppressed. This is the practical content of approximate pole skipping in the present context~\cite{chen2010real,yang2013branching,yang2013quasinormal,kubota2026poleskipping}. The higher overtones are therefore not absent or exactly cancelled; rather, their residues can become sufficiently small that filtering them provides little additional information and may instead enhance filter-induced structure or numerical contamination.

For moderately high remnant spins, the first effective pole--zero pairing occurs near the third overtone. This provides a physical motivation for removing only the fundamental mode and the first two overtones: these filters remove the dominant low-order resonant structure, while the independent imprint of the higher overtones is already partially screened by nearby Matsubara zeros. The appropriate choice of overtones remains spin dependent, as is evident from the breakdown of this pairing toward the Schwarzschild limit.

Thus, for moderately high spins, at positive, real-valued frequencies, or $\omega> 0$, we have an approximate inhomogeneous representation of the waveform  
\begin{equation}
    \widetilde{\psi}_4 =\frac{\widetilde{\mathcal{I}}}{i\omega B^{\rm in}} \sim  \frac{\prod\limits_{j=1,2}(\omega - \omega_{\rm MB}^{(j)}) ({\rm trajectory\,integral})}{\prod\limits_{0\le n\le 2} (\omega-\omega_{\ell mn})} .
\end{equation}
Once the $n=0,1,2$ QNMs have been filtered out, the dynamics of the trajectory are revealed, subject to screening by the first two Matsubara factors. Although the higher QNM overtones with $n\geq3$ do not appear explicitly in this inhomogeneous representation, the trajectory integral can still contain contributions at those corresponding frequencies. 

The above grouping of zeros and poles breaks down at low spins.  In particular, in the Schwarzschild case, the corresponding QNM zeros and Matsubara poles of $B^{\rm in}$ remain separated by a finite distance.  This may contribute to the absence of horizon frequencies in the Schwarzschild extreme-mass-ratio inspirals.  It provides an interesting example in which the Schwarzschild limit is not qualitatively representative of the general Kerr case. 

\subsection{Source driving near a response zero}
\label{app:source_pole_screening}

Because we filter out only the QNMs with $n\leq2$, one may ask whether the resulting signal simply consists of the remaining $n\geq3$ overtones. If the late-time, horizon-controlled trajectory is characterized by a Matsubara frequency, i.e., a source pole coincides with a Matsubara zero of the relevant black-hole response, pole--zero cancellation can reorganize the source-driven contribution into the neighboring higher-overtone QNM poles. The filtered signal may therefore exhibit higher-overtone-like QNM frequency content, but its excitation remains tied to the source dynamics rather than representing an independently added free ringdown. This mechanism is analogous to {\it antiresonant driving} in the complex-frequency plane~\cite{Fano:1961zz,Miroshnichenko:2010fano,weller2026}. By contrast, a source frequency not matched to a Matsubara zero would generically leave an additional driven contribution in the waveform. We discuss these points in detail below.

The same pole--zero structure of the Kerr response, namely the proximity of the Matsubara frequencies to higher QNM overtones, provides a complementary, albeit schematic, interpretation of how source-driven information can be reorganized into a QNM-like response. Consider the local behavior of the response function at a frequency $\omega$ near a QNM pole $\omega_Q\equiv\omega_{\ell mn}$ and a zero $\omega_{\rm MB}^{(j)}$:
\begin{equation}
    \widetilde{\mathcal{G}}_{\rm BH}^{\rm loc}(\omega)
    =
    A(\omega)
    \frac{\omega-\omega_{\rm MB}^{(j)}}
         {\omega-\omega_Q}.
    \label{eq:local_response_source}
\end{equation}
Suppose that the effective source contributes a simple pole at the complex frequency $\omega_{\rm d}$,
\begin{equation}
    \widetilde{\mathcal{I}}^{\rm loc}(\omega)
    =
    \frac{C}{\omega-\omega_{\rm d}}.
    \label{eq:local_driven_source}
\end{equation}
Treating $A(\omega)$ as slowly varying, the corresponding waveform is
locally
\begin{align}
    \widetilde{\psi}_4^{\rm loc}(\omega)
    &\simeq
    A C
    \frac{\omega-\omega_{\rm MB}^{(j)}}
         {(\omega-\omega_Q)(\omega-\omega_{\rm d})}
    \nonumber\\
    &=
    A C
    \left[
    \frac{\omega_Q-\omega_{\rm MB}^{(j)}}
         {\omega_Q-\omega_{\rm d}}
    \frac{1}{\omega-\omega_Q}
    +
    \frac{\omega_{\rm d}-\omega_{\rm MB}^{(j)}}
         {\omega_{\rm d}-\omega_Q}
    \frac{1}{\omega-\omega_{\rm d}}
    \right].
    \label{eq:local_source_response}
\end{align}
If the driving frequency coincides with the response zero, $\omega_{\rm d}=\omega_{\rm MB}^{(j)}$, the explicitly driven pole cancels and Eq.~\eqref{eq:local_source_response} reduces to
\begin{equation}
    \widetilde{\psi}_4^{\rm loc}(\omega)
    \simeq
    \frac{A C}{\omega-\omega_Q}.
    \label{eq:screened_source_response}
\end{equation}
The source can therefore continue to influence the waveform even though its own complex frequency does not appear as an independent oscillation; instead, its contribution is reorganized into the neighboring natural response of the black hole. This resembles antiresonant, or Fano-like, interference in the complex-frequency plane~\cite{Fano:1961zz,Miroshnichenko:2010fano}, although it need not produce a conventional real-frequency Fano line shape.

Conversely, if $\omega_{\rm d}$ is unrelated to the Matsubara structure, the second term in Eq.~\eqref{eq:local_source_response} generally
survives. Near a response zero, its amplitude scales as
\begin{equation}
    C_{\rm d}
    \propto
    \omega_{\rm d}-\omega_{\rm MB}^{(j)}.
    \label{eq:driven_residual_scaling}
\end{equation}
The off-Matsubara source frequency at $\omega_d$ would therefore generically leave an additional driven contribution. When $\omega_{\rm d}$ lies farther from $\omega_{\rm MB}^{(j)}$ than the neighboring QNM frequency $\omega_Q$, the driven contribution at $\omega_{\rm d}$ will generally be larger than the QNM contribution at $\omega_Q$. More specifically, their absolute amplitude ratio is 
\begin{equation}
    \frac{|\mbox{amplitude at $\omega_d$}|}{|\mbox{amplitude at $\omega_Q$}|}
    \simeq \left|\frac{\omega_{\rm d}-\omega_{\rm MB}^{(j)}}{\omega_{\rm Q}-\omega_{\rm MB}^{(j)}}\right|.
\end{equation}
In the Kerr case, the absence of an uncancelled non-QNM frequency, together with the recovery of a consistent trajectory using different filtering prescriptions, can nevertheless provide a nontrivial consistency condition on the Kerr source-response relation.

Taken together, the above two cases show that, if $\omega_{\rm d}$ is varied on scales larger than $|\omega_Q-\omega_{\rm MB}^{(j)}|$, the response in Eq.~\eqref{eq:local_source_response} predicts outgoing radiation with a complex frequency approximately tracking $\omega_{\rm d}$, regardless of whether $\omega_{\rm d}$ lies near or far from $\omega_{\rm MB}^{(j)}$. This is consistent with the picture developed in the previous subsection. In particular, this provides evidence that the QNM-like wave when $\omega_{\rm d}=\omega_{\rm MB}^{(j)}$ is driven instead of free. 

\section{Detailed complex-frequency evolution tracks for NR and CCE waveforms}
\label{app:detailed_tracks}

In Figs.~\ref{fig:SXS_KM_1}--\ref{fig:SXS_CCE_2}, we present the complex-frequency evolution of the QNM-filtered $(2,2)$ spherical-harmonic component of the Bondi news for a set of NR and CCE waveforms. 
As a time derivative of the strain, the Bondi news suppresses constant offsets, secular drifts, and very-low-frequency artifacts while retaining the same underlying oscillatory content as $\psi_4$, up to an additional time derivative.
These results are discussed in Sec.~\ref{subsec:nr_cce}.

\renewcommand{\thefigure}{B\arabic{figure}}
\setcounter{figure}{0}

\begin{figure*}
    \includegraphics[width=\textwidth]{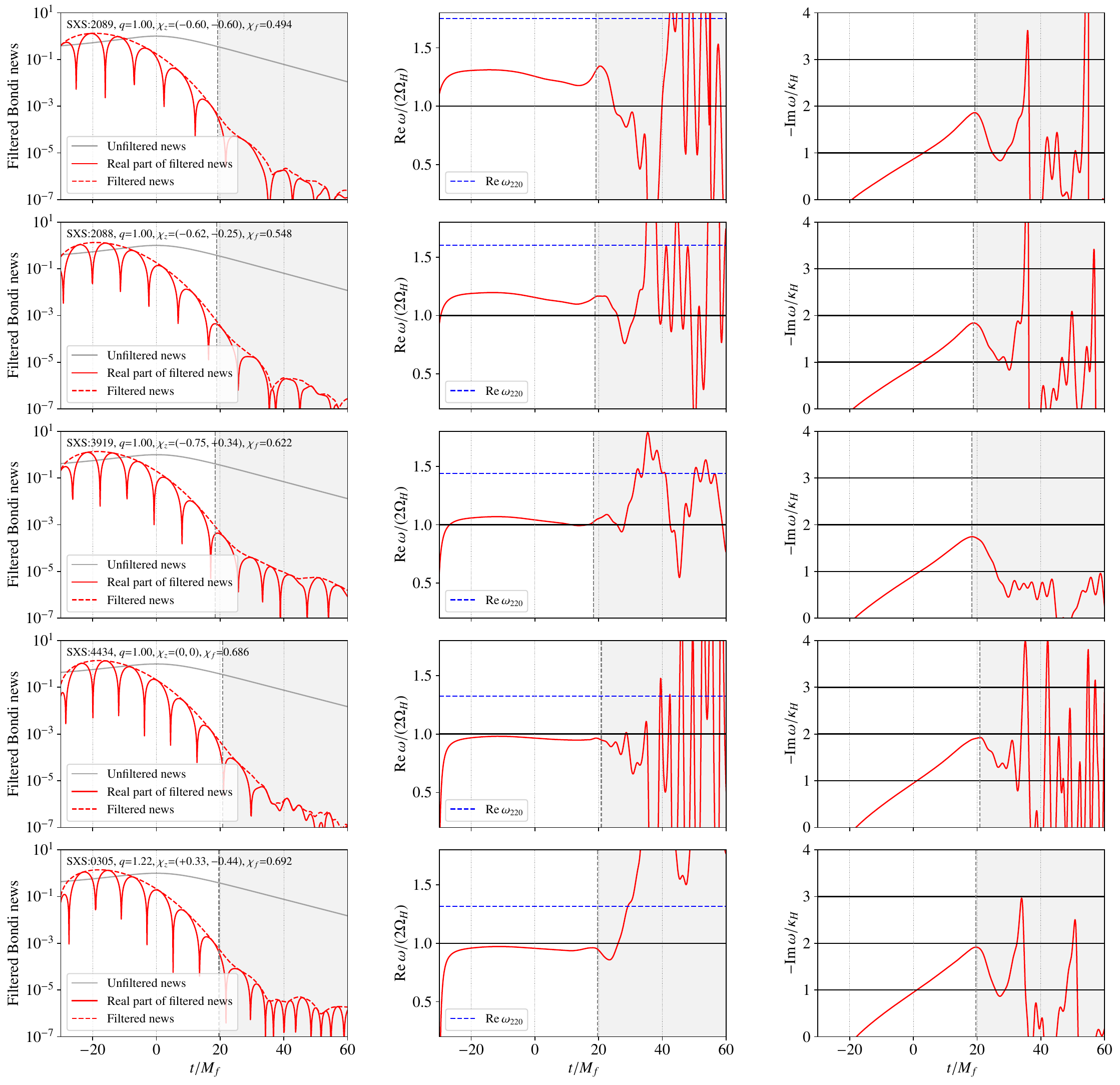}
    \caption{NR waveforms from the SXS catalog studied in Ref.~\cite{kankani2026}. The first column shows the unfiltered (gray) and filtered Bondi news (red), with QNMs up to $\ell=6$ and $n = 2 $ removed. The second and third columns show $\mathrm{Re}\,\omega_{\rm DW}/(2\Omega_H)$ and $-\mathrm{Im}\,\omega_{\rm DW}/\kappa_H$, respectively (red). We set $t=0$ at the peak of the unfiltered $(2,2)$ Bondi news and shift the filtered waveform by the filter's zero-frequency group delay to align its inspiral portion with the unfiltered signal. The vertical dashed line marks the empirically defined transition time, taken to be the first local maximum of $|\mathrm{Im} \omega_{\rm DW}|$, and the shaded region indicates the subsequent evolution, where the filtered waveform undergoes a qualitative change and the direct-wave extraction is no longer reliable. At sufficiently late times, the filtered waveform becomes dominated by numerical error and residual non-QNM components, producing a noise floor in which the instantaneous complex frequency is no longer meaningful. For lower remnant spins, $\mathrm{Re}\,\omega_{\rm DW}$ initially lies above $2\Omega_H$ and gradually decreases toward it, while the decay rate $-\mathrm{Im}\,\omega_{\rm DW}$ increases  to values of order $2\kappa_H$ before numerical contamination dominates. Its subsequent decrease occurs within the shaded region and coincides with the onset of the late-time noise floor; it should therefore not be interpreted as evidence that the physical direct wave begins to decay more slowly.}
    \label{fig:SXS_KM_1}
\end{figure*}

\begin{figure*}
    \includegraphics[width=\textwidth]{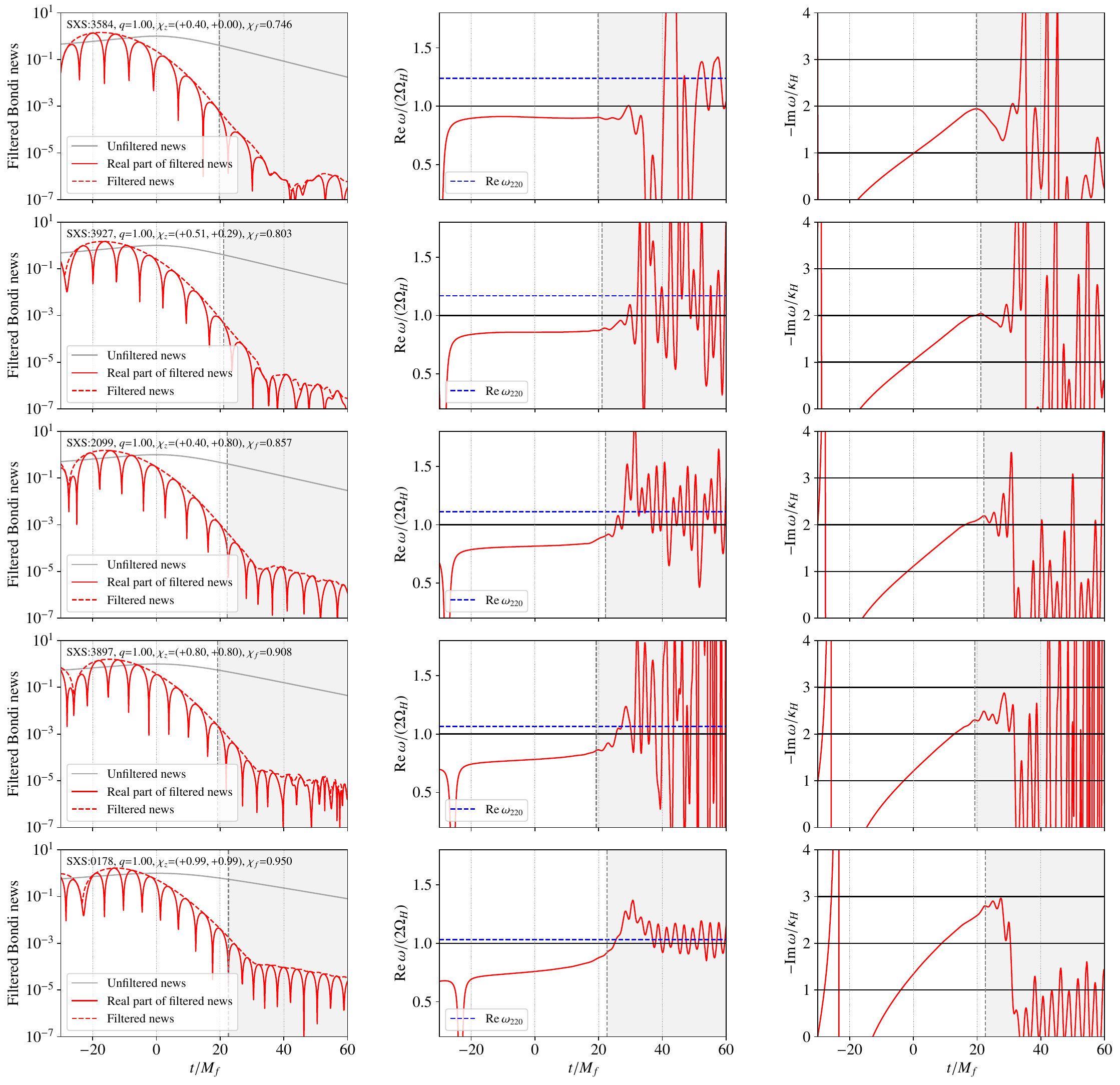}
    \caption{Continuation of Fig.~\ref{fig:SXS_KM_1}. For higher remnant spins, $\mathrm{Re}\,\omega_{\rm DW}$ initially lies below $2\Omega_H$ and gradually increases toward it, while the decay rate $-\mathrm{Im}\,\omega_{\rm DW}$ increases through $2\kappa_H$, reaching values near $3\kappa_H$ in some cases.}
    \label{fig:SXS_KM_2}
\end{figure*}

\begin{figure*}
    \includegraphics[width=\textwidth]{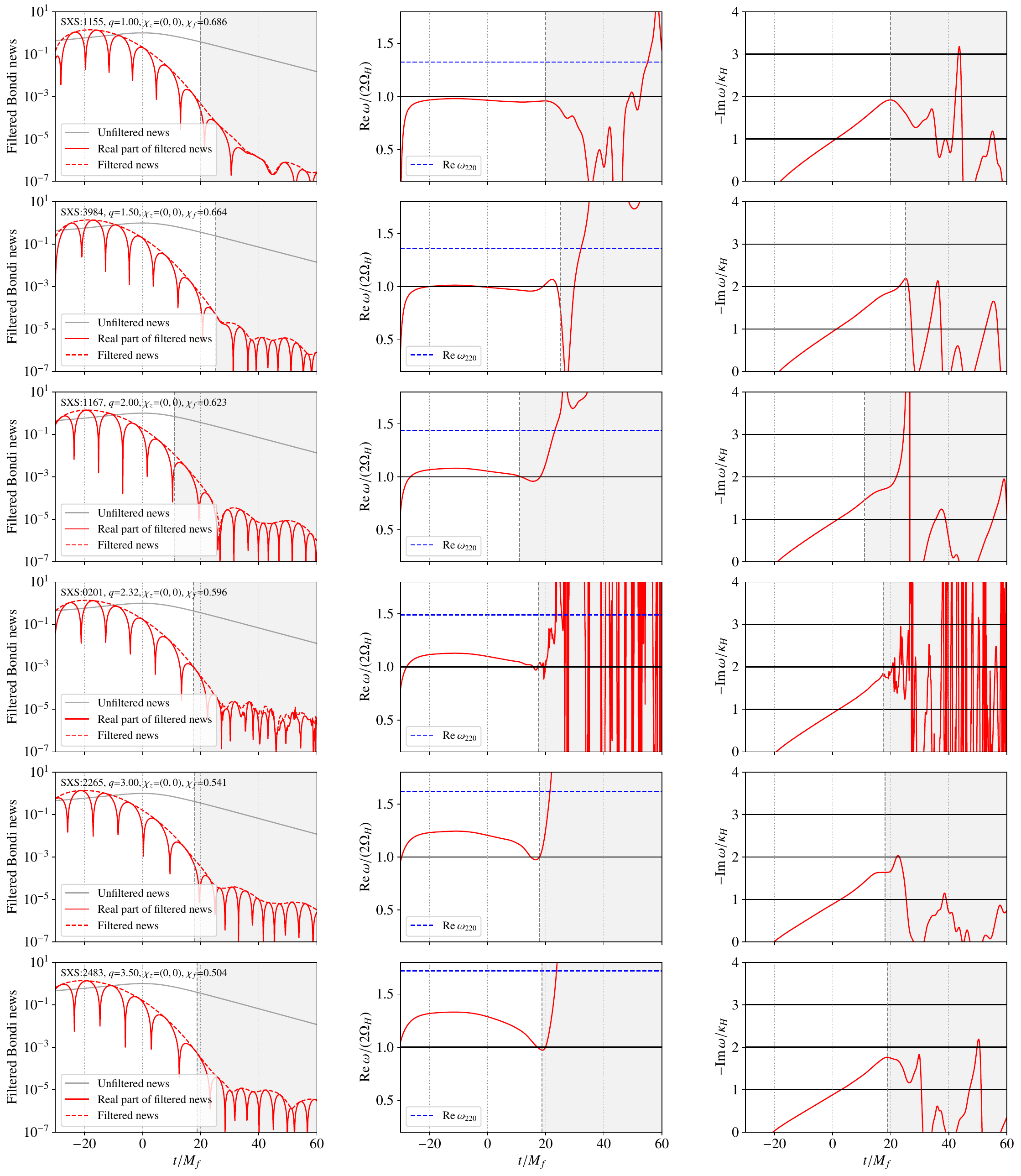}
    \caption{Similar to Figs.~\ref{fig:SXS_KM_1}--\ref{fig:SXS_KM_2}, but for nonspinning binaries with mass ratios ranging from $q=1$ to $3.5$, corresponding to final black-hole spins from $\chi_f\simeq0.68$ down to $0.50$. The same trends observed in the preceding two figures are evident here. Since the progenitor black holes are nonspinning, the final spin in this set does not exceed $\chi_f\simeq0.68$.}
    \label{fig:SXS_nonspinning}
\end{figure*}

\begin{figure*}
    \includegraphics[width=\textwidth]{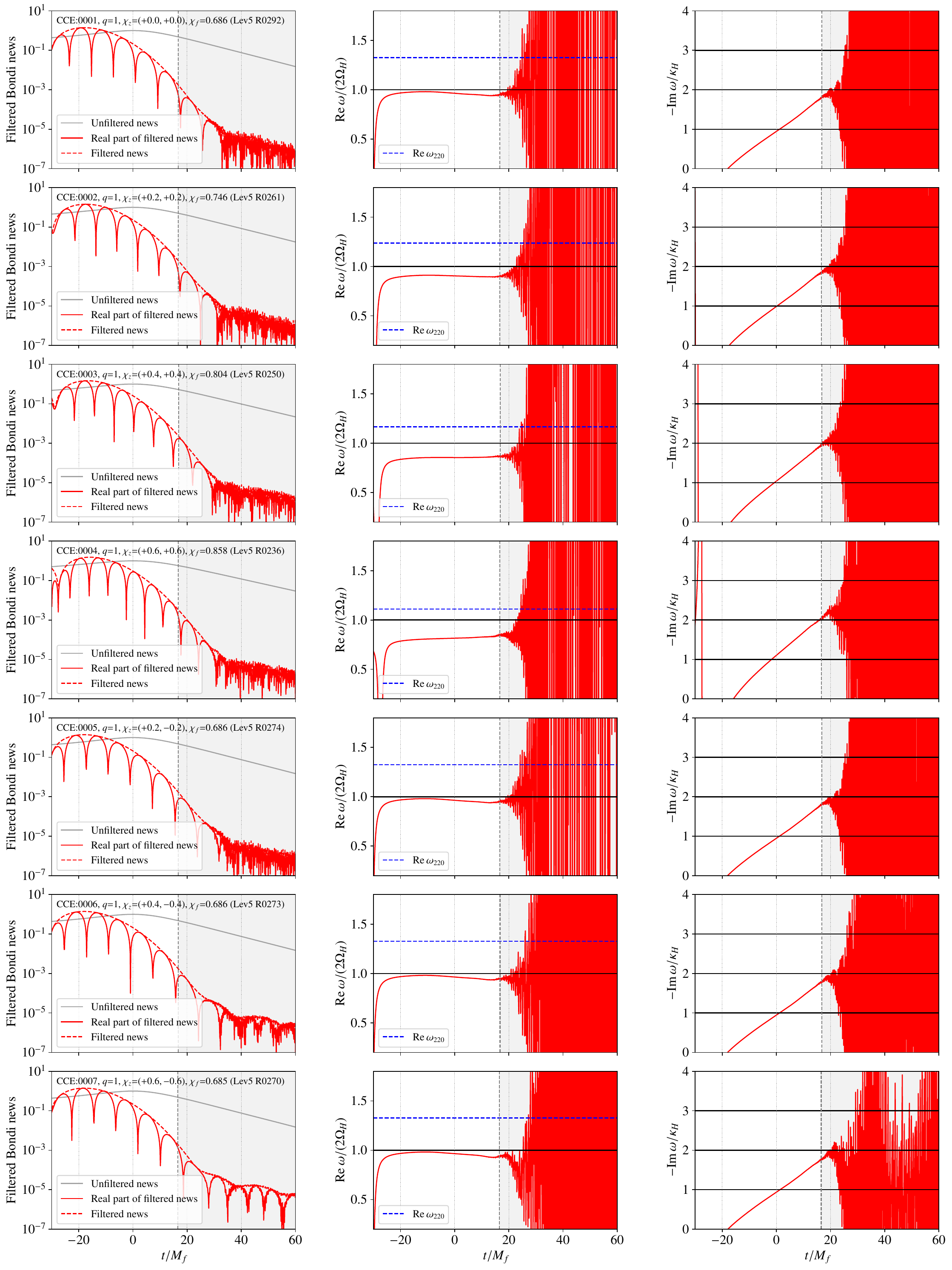}
    \caption{Similar to Figs.~\ref{fig:SXS_KM_1}--\ref{fig:SXS_KM_2}, but for CCE waveforms of spin-aligned binaries with $q=1$. \label{fig:SXS_CCE_1}}
\end{figure*}

\begin{figure*}
    \includegraphics[width=\textwidth]{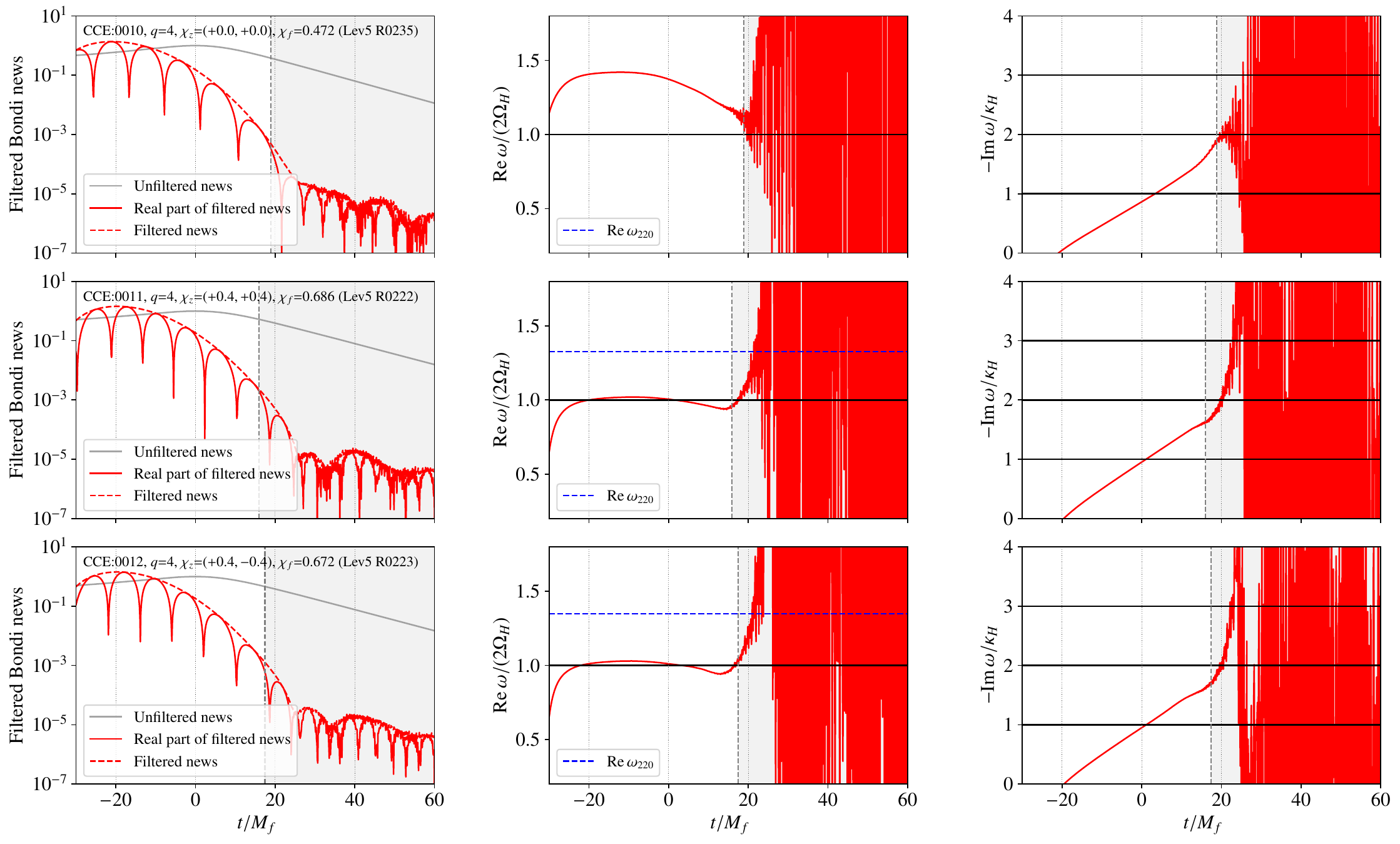}
    \caption{Similar to Figs.~\ref{fig:SXS_KM_1}--\ref{fig:SXS_KM_2}, but for CCE waveforms of spin-aligned binaries with $q=4$. \label{fig:SXS_CCE_2}}
\end{figure*}

\bibliography{refs}

\end{document}